\documentclass[useAMS,usenatbib]{mn2e}          %mn2e (May 2002)

\usepackage{graphicx}
\usepackage{times}
\usepackage[usenames, dvipsnames]{xcolor}
\usepackage{rotating}
\usepackage{txfonts}
\usepackage{longtable,lscape}
\usepackage{anysize}
\usepackage{amssymb}
\usepackage{natbib}
\usepackage[hidelinks]{hyperref}
\hypersetup{colorlinks=false}

\def\be{\begin{equation}}
\def\ee{\end{equation}}
\def\kms{{\rm\,km\,s^{-1}}}

\def\pc{{\rm\,pc}}
\def\Msun{{\rm\,M_\odot}}

\def\deg{{^\circ}}
\def\sqdeg{${\rm deg}^2$}
\def\kpc{{\rm\,kpc}}
\def\masyr{{\rm\,mas/yr}}
\def \sun{{_\odot}}
\def\1s{{1$\sigma$}}
\def\2s{{2$\sigma$}}
\def\3s{{3$\sigma$}}
\def \Lsun {\rm\,L\sun}

\def\uf{UFDG}
\def\ufs{UFDGs}

\def\gaia{\emph{Gaia }}

\newcommand{\appropto}{\mathrel{\vcenter{
  \offinterkip\halign{\hfil$##$\cr
    \propto\cr\noalign{\kern2pt}\sim\cr\noalign{\kern-2pt}}}}}

\title[Detection of \ufs~with Gaia]{Detection of satellite remnants in the Galactic Halo with \gaia -- III.
Detection limits for Ultra Faint Dwarf Galaxies}     
\author[T. Antoja, C. Mateu,  L. A. Aguilar, et al. ] {T. Antoja$^{1}$\thanks{E-mail:tantoja@cosmos.esa.int}\thanks{ESA Research Fellow.}, C. Mateu$^{2,3}$, L. Aguilar$^{2}$,  F. Figueras$^{4}$, E. Antiche$^{4}$,  F. Hern\'andez-P\'erez$^{3}$, \newauthor A.G.A. Brown$^{6}$, O. Valenzuela$^{7}$, A. Aparicio$^{5}$, S. Hidalgo$^{5}$, H. Vel\'azquez$^{2}$\\
$^{1}${{Scientific Support Office, Directorate of Science and Robotic Exploration, European Space
Research and Technology Centre (ESA/ESTEC),}}\\
{{Keplerlaan 1, 2201 AZ Noordwijk, The
Netherlands}}\\
$^{2}${{Instituto de Astronomia, UNAM, Apartado Postal 877, 22860 Ensenada, B.C., Mexico}}\\
$^{3}${{Centro de Investigaciones de Astronom\'{\i}a, AP 264, M\'erida 5101--A, Venezuela}}\\
$^{4}${{Departament d'Astronomia i Meteorologia, Institut de Ci\`encies del Cosmos, Universitat de Barcelona, IEEC, Mart\'{\i} Franqu\`es 1, E--08028,}}\\{{ Barcelona, Spain}}\\
$^{5}${{Instituto de Astrof\'{\i}sica de Canarias, Tenerife, Spain}}\\
$^{6}${{Sterrewacht Leiden, Leiden University, PO Box 9513, 2300 RA Leiden, the Netherlands}}\\
$^{7}${{Instituto de Astronomia, UNAM, Apartado Postal 70--264, 04510 M\'exico D.F., M\'exico}}
}

\begin{document}

\date{Accepted. Received ; in original form }
\pagerange{\pageref{firstpage}--\pageref{lastpage}} \pubyear{2007}
\maketitle
\label{firstpage}

\begin{abstract}
 We present a method to identify Ultra Faint Dwarf Galaxies (\ufs) candidates in the halo of the Milky Way using the future \gaia catalogue and we explore its detection limits and completeness. The method is based on the Wavelet Transform and searches for over-densities in the combined space of sky coordinates and proper motions, using kinematics in the search for the first time. We test the method with a \gaia mock catalogue that has the Gaia Universe Model Snapshot (GUMS) as a background, and use a library of around 30\,000 \ufs~simulated as Plummer spheres with a single stellar population. For the \ufs~we use a wide range of structural and orbital parameters that go beyond the range spanned by real systems, where some \ufs~may remain undetected. We characterize the detection limits as function of the number of observable stars by \gaia in the \ufs~with respect to that of the background and their apparent sizes in the sky and proper motion planes.
 We find that the addition of proper motions in the search improves considerably the detections compared to a photometric survey at the same magnitude limit. Our experiments suggest that \gaia will be able to detect \ufs~that are similar to some of the known \ufs~even if the limit of \gaia is around 2 magnitudes brighter than that of SDSS, with the advantage of having a full-sky catalogue. We also see that \gaia could even find some \ufs~that have lower surface brightness than the SDSS limit. 
\end{abstract}

\begin{keywords}
The Galaxy: halo, formation - galaxies: dwarf - dark matter - methods: data analysis - Astronomy \& Celestial Mechanics: astrometry
\end{keywords}

%==========================================================================================================

\section{Introduction}
\label{sec_intro}

The current cosmological cold dark matter paradigm posits the assemblage of large structures in the Universe from smaller ones \citep{Press1974, White1978, Springel2006}. A galaxy like ours must have formed by the merger of a large number of smaller systems, that even today, must be still in the process of being accreted. 
A discrepancy between the predicted and observed number of galaxy satellites has given rise to the so called ``missing satellite problem'' \citep{Klypin1999,Moore1999}. However, in recent years, an entirely new population of hitherto unknown systems with very low luminosity and surface brightness, dominated by dark matter, the so called ``Ultra Faint Dwarf Galaxies'' (\ufs), has been discovered, opening up the possibility of resolving this problem \citep[e.g.][]{Simon2007,Bullock2010}. The knowledge of their structural properties, chemical abundances and stellar populations is also key to understanding fundamental issues \citep[see review by][]{Belokurov2013} like the process of star formation and the role of feedback in these relatively low-mass environments \citep{Brown2014}; how to distinguish between a dwarf galaxy and a globular cluster in some extreme cases \citep[e.g.  Segue 1 and 2, Willman 1, Boo II and CmB,][]{Forbes2011}; or to what extent UFDGs could have contributed to the stellar population found in the Galactic halo today \citep[e.g.][]{Kirby2008}.

So far, all known \ufs~were discovered as over-densities in deep large-area photometric surveys, the vast majority in SDSS \citep[e.g.][]{Willman2005b,Willman2005a,Zucker2006a,Zucker2006b,Belokurov2006,Belokurov2007,Belokurov2008,Belokurov2009,Belokurov2010}, together with recent findings in Pan-STARRS \citep{Laevens2015} and the Dark Energy Survey \citep[DES,][]{Koposov2015, Bechtol2015}.

The ESA \gaia mission, launched in December 2013, offers excellent prospects for the discovery of new members of the \uf~population. \gaia \citep{deBruijne2012, Perryman2001} will measure accurate positions, parallaxes and proper motions for all stars out to its survey limit of $G=20$ ($V=20$--$22$, depending on the colour of the source), where $G$ is the white light photometric pass-band of \gaia \citep{Jordi2010}. Multi-colour photometry will be obtained for all stars and radial velocities will be collected for stars at $G_\mathrm{RVS}<16$ mag, where $G_\mathrm{RVS}$ indicates the pass-band of the Radial Velocity Spectrograph on-board. \gaia will also provide astrophysical information on all the sources observed, primarily through multi-colour photometry. The astrophysical parameters of all \gaia sources will be provided as part of the survey data products \citep{BailerJones13}.

Although the \gaia survey is not as deep as SDSS, Pan-STARRS or DES, it is all sky at a spatial resolution comparable to that of Hubble Space Telescope ({\em HST}), and will deliver high accuracy astrometry (positions and proper motions) for all sources. The combination of these unique features is what makes the comparatively shallow survey of \gaia potentially powerful in the search for \ufs. Here we aim to exploit this in a technique to identify \ufs. The combination of positions and kinematics has proven to be most efficient in the search for dark matter subhalos in cosmological simulations (e.g. \citealt{Behroozi2013,Onions2012}). But, to our knowledge, this is the first time that both configuration space and kinematics are included in the search of \ufs. As we will show, \gaia will enable  us to probe parts of the \uf~parameter space which have not been covered before, and will allow for a comprehensive study of the spatial distribution around the Milky Way (MW) of this faint galaxy population.

The present work continues the series \citep{Brown2005, Mateu2011}, in which we have assumed the task of building ever more realistic \gaia mock catalogues, and used them to test tools that we have introduced to detect and characterize substructure in the stellar halo of our Galaxy.

In Section 2 we introduce our \gaia mock catalogue, which serves as our laboratory to study the detectability of \ufs. This includes a stellar background and our synthetic \ufs. The details of the \gaia selection function and error model used to generate the \gaia observables are described as well. In Section 3, we present our detection tool, which consists of a peak identifier that is applied in the sky and proper motion planes, a cross-matcher that identifies peaks with common members in both planes, and a procedure to evaluate the statistical significance of the matched peaks. Section 4 presents our results. Detection limits are shown as a function of astrophysical parameters and of ``effective parameters'', namely a combination of the former on which our detection method depends directly. In Section 5 we summarize the limits of our method as well as the assumptions that it is based on. Our conclusions are presented in Section 6.

%-==========================================================================================================

\section{The {\em Gaia} mock catalogue}
\label{sec_Database}

Our \gaia mock catalogue is the stage where we assess the success, efficiency and limits of our \ufs~detection technique. As such, it represents a controlled, but realistic environment. There are several elements that compose it. First, we need a model of the Galaxy, from which suitable stellar backgrounds can be extracted (Section~\ref{subsec_backg}). We also need a mass model for our synthetic \ufs~and a stellar population model (Section~\ref{subsec_UFGXs}). The latter is because our \ufs~are not merely ensembles of particles, but stellar properties must be assigned  to them, as they impinge on the value and quality of their \gaia observables. 
The previous elements allow us to assemble an extensive library of \ufs~at various distances and with a wide range of intrinsic parameters, projected against stellar backgrounds at several Galactic latitudes. We then use a \gaia selection function and error model to transform the theoretical quantities into realistic \gaia observables, as our detection method should work based on them only (Section~\ref{subsec_GaiaMod}). In Section~\ref{subsec_filter} we present the filtering method that we use to eliminate foreground stars. Finally, we examine the nature of the \ufs~projections in the sky and proper motion planes, as these are the basic input variables that our method works on (Section~\ref{subsec_planes}).

%----------------------------------------------------------------------------------------------

\subsection{The Galactic Background Model}
\label{subsec_backg}

We use as a Galactic background the Gaia Universe Model Snapshot (GUMS) from \citet{Robin2012}, which is a simulated catalogue of the sources expected to be observed by {\em Gaia}, at a fixed epoch. It includes the simulation of Galactic sources, Solar system and extragalactic objects.

We note here that \gaia will observed large numbers (potentially millions) of galaxies and about half a million QSOs \citep{deBruijne2015b, DeSouza2014}, which will all appear as faint point sources and could thus complicate the search for \ufs. However, discrete source classification will be part of the published data \citep{BailerJones13} and in this work we assume that we can rely on this to filter out galaxies and QSOs (but see Sect.~\ref{sec_caveats}). Therefore, we use only Galactic sources, and restrict the catalogue to a range in latitude of $20\deg< |b|<90\deg$, to avoid the crowding and high extinction expected near the Galactic plane.

The Galactic sources in GUMS are generated based on the Besan\c{c}on Galactic Model, which includes the Galactic Thin and Thick Disks, Bulge and Halo, based on appropriate density laws, kinematics, star formation histories, enrichment laws, initial mass function (IMF) and total luminosities for each of the component populations, described in detail in \citet{Robin2012}. 
Objects are simulated with masses down to the hydrogen burning limit, corresponding to spectral types down to $\sim L5$. 
Binary and multiple star systems are also simulated \citep[see details in][]{Arenou2011}, introduced with a probability that depends on the mass and evolutionary state of the primary star. The probability distribution for the separations is assumed to be a log-normal with the parameters reported by \citet{Duquennoy1991} for (primary) stars down to solar masses, and from \citet{Close2003} for low mass stars.

%------------------------------------------------------------------------------------------------------------

\subsection{The \uf~Model}
\label{subsec_UFGXs}

%--------------------

\subsubsection{The Dynamical Model}
\label{subsubsec_Dyn}

For our basic synthetic \uf~dynamical model we use a simple Plummer sphere with isotropic velocity distribution. A particular realization of this model is uniquely defined by two of the following parameters:

\begin{itemize}
\item $M_{\rm T}$:  Total mass 
\item $r_o$:   Core radius 
\item $r_{\rm h}$:  Half--mass radius
\item $\sigma_V$:  Velocity dispersion (3-D)
\end{itemize}

The total gravitational binding energy of a Plummer sphere is:

\[
W_P = -f\,{{M_{\rm T}^2}\over{r_o}},\quad {\rm where}\quad f\equiv {{3\pi G}\over{32}}.
\] 
\noindent By virial equilibrium, we can establish a relation between the total mass, core radius and velocity dispersion:
\[
2K = -W\quad \Longrightarrow\quad \sigma_V^2 = {{fM_{\rm T}}\over{r_o}}.
\]
In appropriate astronomical units, the previous relation is:
\[
\left({\sigma_V\over{\kms}}\right) = 0.03559\,\sqrt{{(M_{\rm T}/\Msun)}\over{(r_o/\pc)}}.
\]

Observationally, the scale--length usually reported is the half--light radius. Under the assumption of a position independent mass--to--light ratio (i.e. well mixed), the half--mass radius and its light counterpart coincide. We assume this and use $r_{\rm h}$ indistinctly as the half--light, or the half--mass radius. For a Plummer sphere, the relation between the core and half--mass radius is: $r_{\rm h} = 1.30477\, r_o$.
Then, the previous relation between velocity dispersion, mass and radius can be written as
\begin{equation}
\centering
\left({\sigma_V\over{\kms}}\right) = 0.04066\,\sqrt{{(M_{\rm T}/\Msun)}\over{(r_{\rm h}/\pc)}}.\label{e:plummer_Msrh}
\end{equation}

For a model with a given $r_{\rm h}$ and  $\sigma_V$,  the mass-to-light ratio $M/L$ of the \uf~is given by the ratio of its total mass $M_{\rm T}$, derived from Eq. \ref{e:plummer_Msrh}, and the chosen total V-band stellar luminosity $L_V$. The number of particles in the realization ($N_{\rm s}$) and the total stellar mass ($M_{\rm s}$) are a consequence of the assumed total luminosity, star formation history and stellar mass function.

%-------------------------

\subsubsection{The stellar population model}
\label{subsubsec_StellPop}

  \begin{figure} 
 \centering 
 \includegraphics[width=0.45\textwidth]{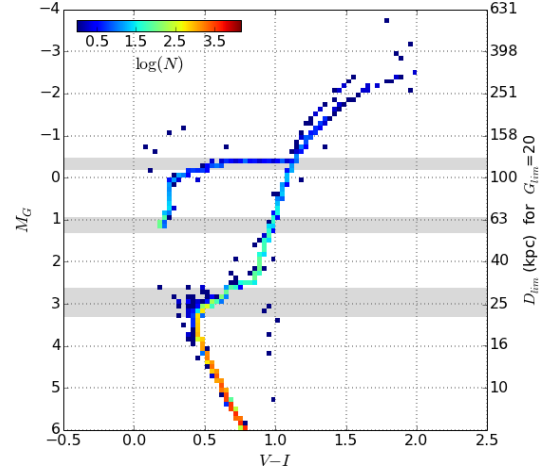} 
 \caption{Hess diagram $M_G$ vs intrinsic $V-I$ colour for the \uf's stellar population model  (see text for details). The colour scale is proportional to the logarithmic number of stars in each bin. The right $y$-axis indicates the maximum distance up to which a star with a given $M_G$ will be observable by {\em Gaia}, given the expected magnitude limit of $G_{lim}=20$ (assuming $A_V=0$). For the gray bands see discussion at end the of Section~\ref{subsec_GaiaMod}}\label{f_cmd_bassic}
 \end{figure}

We simulate the stellar population of the \ufs~as a single star formation burst with an age of $12$ Gyr and metallicity $Z=0.0001$, assuming a \citet{Chabrier2003} IMF. We use the HB13 Stellar Population Synthesis code from \citet{Hernandez-Perez2013}, which allows for a consistent treatment of isolated and binary stars. The prescriptions assumed in this code are similar, yet not identical, to those used for the statistical orbital properties of binaries in GUMS  (see Section \ref{subsec_backg}).
In HB13, the binary probabilities and orbital parameters are randomly drawn and assigned to each primary star in the population at age zero and the evolution is followed using the \citet{Hurley2002} binary evolution code. Binary probabilities are assumed to depend on the mass of the primary using the prescription from \citet{Lada2006}, and the distribution of periods, and thus, separations, of \citet{Duquennoy1991}. The resulting $M_G$ versus intrinsic $V-I$ colour Hess diagram for the stellar population used for the \ufs~is shown in Fig. \ref{f_cmd_bassic}.

%--------------------

\subsubsection{Parameters of the simulated \ufs}
\label{subsec_library}

Each of the simulated \uf~has 9 free parameters:
\begin{enumerate}
\item intrinsic parameters:
  \begin{itemize}
       \item Total V-band luminosity: $L_V$
       \item Half--light radius: $r_{\rm h}$
       \item Velocity dispersion: $\sigma_V$
   \end{itemize}
\item extrinsic parameters:
  \begin{itemize}
        \item Heliocentric distance: D
        \item Position in the sky ($l$, $b$) 
        \item Galactocentric velocity vector modulus: $V_{\rm gal}$
        \item Azimuthal and latitudinal orientation angles of the galactocentric velocity vector: $\phi_V$, $\theta_V$
   \end{itemize}
\end{enumerate}

\begin{table*}
 \caption{Ranges of parameters in the simulated \ufs~(first row) and parameters of the \uf~used as our fiducial case. See Sections~\ref{subsec_library} and \ref{subsec_GaiaMod} for definitions. $N_{\rm obs}$ and $M/L$ can be obtained from the other parameters.}
 \label{t:example}
\centering
\tabcolsep=0.11cm
\begin{tabular}{lcccccccrrccc}\hline\hline
 &        $L_V$               &$r_{\rm h}$      & $\sigma_V$ & $D$     & $l$      & $b$     &$V_{\rm gal}$ &$\phi_V$ &$\theta_V$ &$M/L$                & $M_{\rm s}$           & $N_{\rm obs}$\\
 &       ($\Lsun$)            &($\pc$)    &($\kms$)    &($\kpc$) &($\deg$)  &($\deg$) &($\kms$)  &($\deg$) &($\deg$)   &($\Msun/\Lsun$)      &($\Msun$)           &   \\\hline
ranges & $86.$-$7.5\times10^6$&  5-$4\,000$ &  1-500     & 10-250  &    0-180 &     0-90&  24-550  &  0-360  &$-90$-90   &0.04-$3.9\times10^7$ & 10-$1.6\times10^6$ &10-$10^3$ \\
fiducial&$5\times10^3$        &  80       &  10        & 20      &    90    &       30&      453 &  0      &0          &902                  & $6\times10^3$      &94 \\
\hline
\end{tabular}
\end{table*}

\begin{figure*}  
\centering 
\includegraphics[width=0.85\textwidth]{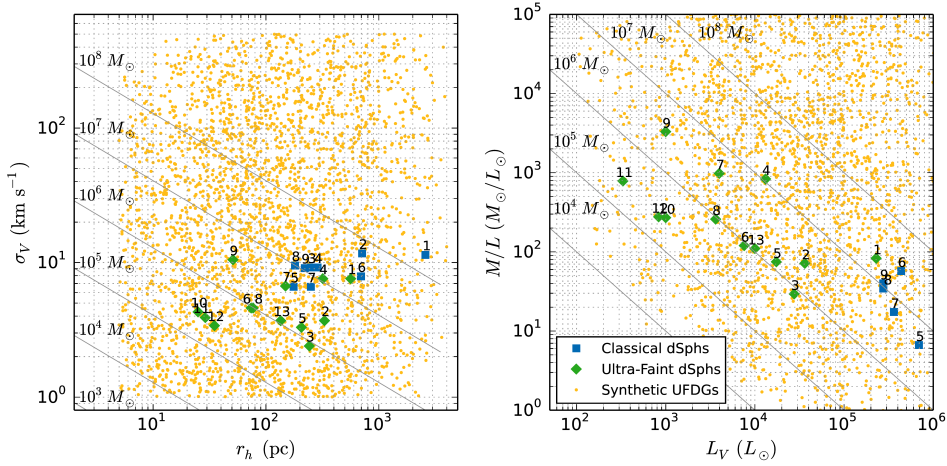} 
\caption{\emph{Left:} $\sigma_V$ vs. $r_{\rm h}$. \emph{Right:} $M/L$ vs. $L_V$. The yellow dots correspond to members of our \uf~library. Note that this library contains only systems with at least 10 stars observable by Gaia. Solid lines indicate constant total-mass models. Known \ufs~and classical dwarf spheroidal galaxies are shown with green diamonds and blue squares respectively \citep[data from][]{McConnachie2012}. The labels correspond to:  Sgr (1), For (2), LeoI (3), Scl (4), LeoII (5), Sex (6), Car (7), UMi (8), Dra (9) for blue squares and CVnI (1), Her (2), Boo (3), UMa (4), LeoIV (5), CVnII (6), UMaII (7), CmB (8), BooII (9), WilI (10), SegI (11), SegII (12), LeoV (13) for green diamonds.  We do not include LeoT, which would not be observable with Gaia, and PscII that lacks measurements on some parameters.  }\label{f_rhsig}
\end{figure*}

We have generated a set of libraries with a total of $\sim$30\,000 \ufs~covering large ranges of the 9 parameters (see Table~\ref{t:example}). Our main library is generated with the following parameters drawn at random: i) the number of stars that would be observable by {\em Gaia} $N_{\rm obs}$, ii) the heliocentric distance $D$, iii) the apparent size of the \ufs~in the sky $\theta$ and iv) in the proper motion plane $\Delta\mu$, and v) the center-of-mass velocity. These quantities are described in detail in Sections~\ref{subsec_GaiaMod} and \ref{subsec_planes}. The first four parameters are generated from a uniform distribution in a logarithmic scale. For the last one, the angles $\phi_V$ and $\theta_V$ and the modulus $V_{\rm gal}$ are generated following a uniform distribution, with $V_{\rm gal}$ between zero and the local escape velocity for the Galaxy\footnote{ We compute the escape velocity as $V_e=V_c \sqrt{2(1-\ln(R_{\rm gal}/r_t))}$, with $V_c=200$ km s$^{-1}$ and $r_t=200$ kpc.}.
The remaining parameters (namely $L_V$, $r_{\rm h}$ and $\sigma_V$) are obtained from the ones above. We also require $N_{\rm obs}$ to be at least 10 in this library. This library is designed with particular goals described in detail in Sections~\ref{subsec_RedParam} and \ref{subsec_limits}. 

%Some of the libraries have all parameters fixed except 2 of them that are varied uniformly in linear or logarithm distribution, to study the detection performance as function of specific parameters. These libraries are used and described in Section~\ref{subsec_PhysParam}.
%Finally, part of this library is replicated with the same structural parameters but located in different directions in the sky. These set of libraries are used in Section~\ref{subsec_bg}.

Fig.~\ref{f_rhsig} illustrates the range explored in half-light radii and velocity dispersion (left panel), as well as in mass--to--light ratio and total V-band luminosity (right). The observed values of these parameters for known \ufs~and classical dSph galaxies are shown with green diamonds and blue squares, respectively. Note that the range explored by our synthetic library (yellow dots) is much larger than the observed one for $r_{\rm h}$, $\sigma_V$ and $M/L$. In particular, the large range covered in $\sigma_V$ results in a very large range of $M/L$ (the library spans an even larger range of $M/L$ than shown in Fig.~\ref{f_rhsig}). We are pushing the limits of the parameter space explored, towards regions where the detection would be observationally more difficult, i.e. towards larger $r_{\rm h}$ and $\sigma_V$ (top and right areas of left panel), and lower luminosity and high $M/L$ (top and left areas of right panel).  
The fact that $N_{\rm obs}$ is generated uniformly, together with the large scatter in luminosity for small $N_{\rm obs}$ due to stochastic effects, produces the diffuse boundary in $L_V$ in the right panel.

%--------------------

\subsection{The \gaia selection function and error model}
\label{subsec_GaiaMod}

Here we present our model for the \gaia observations that includes the selection function and the \gaia error model that we apply to the GUMS model and the simulated \ufs. The \gaia observables are the 5 astrometric parameters ($l$, $b$, $\varpi$, $\mu_l$, $\mu_b$), the radial velocity, the \gaia photometry (including the $G$ \gaia magnitude and the two broad band magnitudes $G_{\rm{BP}}$ and $G_{\rm{RP}}$).
The final \gaia catalogue will also provide three atmospheric parameters (metallicity, surface gravity and effective temperature) and extinction. The true values for these observables and parameters are obtained directly from the models. The conversion from the Johnson-Cousins photometric system to \gaia magnitudes is done following the transformation given in table 3 from \citet{Jordi2010}. We do not consider extinction because all fields used in our study are at relatively high latitudes (at least $30\deg$).

The GUMS model and the simulated \ufs~include binary and multiple systems. To determine which ones will be resolved by {\em Gaia}, we use a prescription used within the Data Processing and Analysis Consortium\footnote{\url{http://www.cosmos.esa.int/web/gaia/dpac} (\citealt{Mignard08})}. In this model the minimum angular separation on the sky that \gaia can resolve depends on the apparent magnitudes of the stars in the system, with the minimum separation being $\sim38$ mas. For the unresolved cases, a single detection is considered by computing the total integrated magnitude, averaging positions and taking the atmospheric parameters (such as surface gravity) of the primary star in the system.

As an example, if we take a field\footnote{
In what follows, we always work with $2\times2\deg$ fields. To cover the same solid angle, regardless of latitude, we have converted the Galactic longitude of the stars $l$, to $l'=(l-l_0)*\cos(b_0)+l_0$, where $l_0$ and $b_0$ are the longitude and latitude of the center of the field, respectively.  For simplicity, we use $l$ instead of $l'$ hereafter. } of $2\times2\deg$ centered at $l=90\deg$ and $b=30\deg$, there are initially 25\,521 objects, from which 57 per cent are single stars, 13 per cent are stars of resolved multiple systems and 30 per cent are unresolved systems\footnote{After the cuts in parallax and surface gravity (see Section~\ref{subsec_filter}), these fractions become 54, 5 and 41 per cent, respectively. The relative increase of unresolved systems is because we are selecting large distances and giant stars (dwarf stars have been removed), which have higher binary fractions.}. For a simulated \uf~at $50$ kpc, these fractions are 63, 6 and 31 per cent, respectively. 

To simulate \gaia-like errors for the GUMS catalogue and the simulated \ufs, we use the code presented in \citet{RomeroGomez15}, updated to the post-launch performance\footnote{The code was released at the {\em $2^{nd}$ Gaia Challenge Workshop} and is publicly available at \url{https://github.com/mromerog/Gaia-errors}}
as described in \cite{deBruijne15}. Up to date information is available from the \gaia web pages\footnote{\url{http://www.cosmos.esa.int/web/gaia/science-performance}}. The uncertainties on the astrometry, photometry and spectroscopy are mainly functions of the magnitude and colour.
The geometrical factors and the effect of the number of passages due to the scanning law are also taken into account\footnote{\url{http://www.cosmos.esa.int/web/gaia/table-6}}. 
%The errors on the atmospheric parameters are computed following \citet{BailerJones13}.
For the surface gravity we take a constant error of 0.25 dex, based on table~4 of \citet{BailerJones13}. 
Lacking a model of the \gaia performances for unresolved systems, we use the same prescriptions as for single or resolved stars. Only stars with magnitude $G<20$ (the \gaia magnitude limit) are considered.

From all the \gaia astrometric observables, we can not make use of parallaxes to infer distances to \ufs~stars. The median relative error in parallax of the stars in the \ufs~in the range of distances considered here is at least of 70 per cent and on average 170 per cent, since they can be very faint and distant objects. Besides, radial velocities are not available for most of the cases as 90 per cent of the \ufs~in the range of distances explored here have at most 10 per cent of stars that are brighter than the magnitude limit of the \gaia spectrograph ($G_{RVS}=16$). Therefore, we use as our observables only the two angular positions in the sky ($l$ and $b$) and the two proper motions ($\mu_{\ell*}\equiv\mu_\ell\cos(b)$ and $\mu_b$). The errors in the angular coordinates in the sky are of the order of 0.05--0.4 $\rm{mas}$ whereas in proper motion these are about 0.03--$0.3~\rm{mas/yr}$.

\begin{figure} 
\centering 
\includegraphics[width=0.38\textwidth]{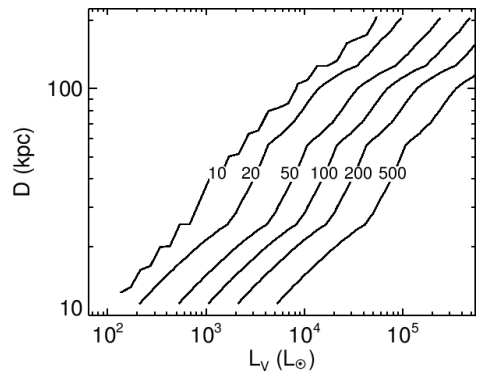} 
\caption{Number of stars observed by \gaia $N_{\rm obs}$, as a function of the total V-band luminosity and distance of the \uf.}
\label{f_Nobs}
\end{figure}

The number of \uf~stars seen by \gaia $N_{\rm obs}$ (not the same as the total number of stars in the realization $N_{\rm s}$) is determined by the total stellar luminosity $L_V$ of the system and the distance of the \uf~(given an assumed stellar population model). On the right axis of Fig.~\ref{f_cmd_bassic} we indicate the distance limit associated to the \gaia deepest magnitude $G=20$. Note that at distances larger than $25\kpc$ only giant stars are observed. In Fig.~\ref{f_Nobs} we show the number of  \uf~stars observed by \gaia as function of luminosity and distance. For instance, \ufs~of luminosity around $1000\Lsun$ have no stars bright enough to be observed by \gaia beyond $\sim40\kpc$ but they will have around 15 observable stars around $23\kpc$. The oscillations with distance present at around 25, 60 and $120\kpc$ are because the type of stars of the \uf~population that \gaia can detect changes as it is observed at different distances,  depending on whether or not features like the main sequence turn-off are observable. These distances have been marked in the Hess diagram of Fig.\ref{f_cmd_bassic} and they correspond to the main sequence turn-off, the extreme horizontal branch and the horizontal branch, respectively. Note also the stochasticity around small numbers of observable stars.

\begin{figure*}  
\centering 
\includegraphics[width=0.85\textwidth]{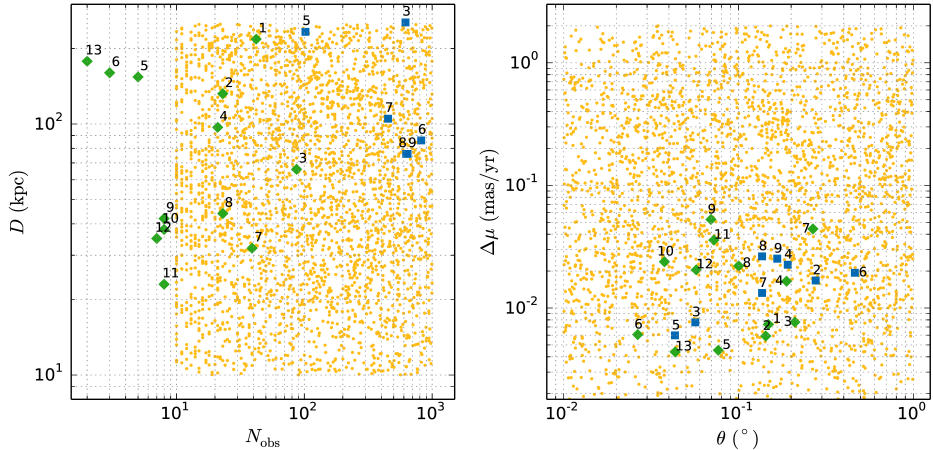}
\caption{\emph{Left:} $D$ vs. $N_{\rm obs}$. 
\emph{Right}: Apparent size $\Delta\mu$ in the proper motion plane vs. apparent size $\theta$ in the sky. The yellow dots show the range spanned by the synthetic \uf~library. Known \ufs~and classical dwarf spheroidal galaxies are shown with green diamonds and blue squares respectively  \citep[data from][]{McConnachie2012}. The labels are as in Fig~\ref{f_rhsig}.}\label{f_dmuth_NobsD}
\end{figure*}

The left panel of Fig.~\ref{f_dmuth_NobsD} shows the range covered by our library in number of stars observable by \gaia $N_{\rm obs}$ and distance $D$ to the Sun. To include the known \ufs~and classical dSphs in this plot we have computed $N_{\rm obs}$ assuming the stellar population model described in Section~\ref{subsubsec_StellPop}, and the total luminosity and distance reported by \citet{McConnachie2012} for these systems\footnote{The SFH assumed in the stellar population model is reasonably representative of the SFH of known \ufs. 
For simplicity we assume the same model for the classical dwarfs to get a rough estimate of $N_{\rm obs}$, although these  have very different SFHs.}.  
Here we can see that there are real systems that go beyond the range covered by our library towards small number of observed stars. We must remember that $N_{\rm obs}$ is the number of stars that would be seen by {\em Gaia}, which has a $\sim2$ mag brighter limit than the SDSS \footnote{The SDSS survey is $2$ mag deeper comparing the $r$ and $G$ bands, or between 1 and $2$ mag deeper comparing the $g$ and $G$ bands. This is estimated by taking the two extreme colours of the stars in our simulated \ufs, that is $V-I=$0.25 and $V-I=$1.5, and convert these to $G-r$ and $G-g$ colours following \citet{Jordi2010}.} used to identify those systems \citep[e.g.][]{Belokurov2007}. 
This is a limitation imposed by \gaia that we cannot get around.
Note also that the boundaries of the regions spanned by the library in this panel are sharp by construction
%, as the library is generated with these parameters drawn at random from a uniform distribution in a logarithmic scale 
(see Section~\ref{subsec_library}). 
%This, together with the large scatter in luminosity for small $N_{\rm obs}$ due to stochastic effects, produces the diffuse boundary in $L_V$ observed Fig.~\ref{f_rhsig} (right panel). 

%--------------------

\subsection{Filtering the foreground}
\label{subsec_filter}

Along a given line-of-sight (LOS), it is important to minimize the number of background stars\footnote{The Galactic sources in GUMS are actually foreground and background stars. We use hereafter ``background'' for simplicity.} $N_{\rm BG}$ with respect to the number of stars in the \uf. We use a parallax cut to filter out foreground disk stars, which have large parallaxes with small errors. Thus, we discard data for stars with $\varpi-e_{\varpi}>0.1$ mas, i.e. an observed parallax which, within the errors, corresponds to distances smaller than $10\kpc$. We also filter out  foreground disc dwarfs with the implementation of a surface gravity $\log g$ cut: we discard stars with $\log g-e_{\log g}>4$, where $\log g$ is the atmospheric parameter derived from the \gaia~observables. With these two cuts, we reduce  $N_{\rm BG}$ typically by an order of magnitude. 
For instance, there were 25\,521 stars in the GUMS model in our fiducial field ($l=90\deg$ and $b=30\deg$) and with the cuts we reduce this number to $N_{\rm BG}$=1\,413.

%With the distance cut we are focusing our search to \uf~that are farther than $10\kpc$. Also, the fraction of \uf~stars that is lost with the cut in $\log g$ can be significant for close systems, as for some of them, part of the dwarf stars are still observed by {\em Gaia}. It is our experience that it is difficult to devise a unique algorithm that can identify our target systems at all distances, and so, the $10\kpc$ limit is introduced as a necessary compromise. It is clear that specifically tailored algorithms could be used for nearby cases. 

We stress that these cuts have been designed to have minimal loss of observable stars from the \ufs, particularly at relatively large distances ($D>10$kpc), for which dwarf stars will not be observable by {\em Gaia}. The fraction of lost stars is up to 70 per cent for nearby 
\ufs~at $10\kpc$. However, it goes down to 30 per cent at $\sim18\kpc$ and is less than 10 per cent for distances larger than $25\kpc$. Nevertheless, with the cuts we are maximizing the relative fraction of \uf~stars with respect to the background in all cases, given that the fraction of stars lost from the background is larger than that of the \ufs.
It is our experience that it is difficult to devise a unique algorithm that can identify our target systems at all distances, and so,  limits are introduced as a necessary compromise. It is clear that specifically tailored algorithms could be used for nearby cases. 

%--------------------

\subsection{The sky and proper motion planes} 
\label{subsec_planes}

\begin{figure} 
\centering 
\includegraphics[width=0.4\textwidth]{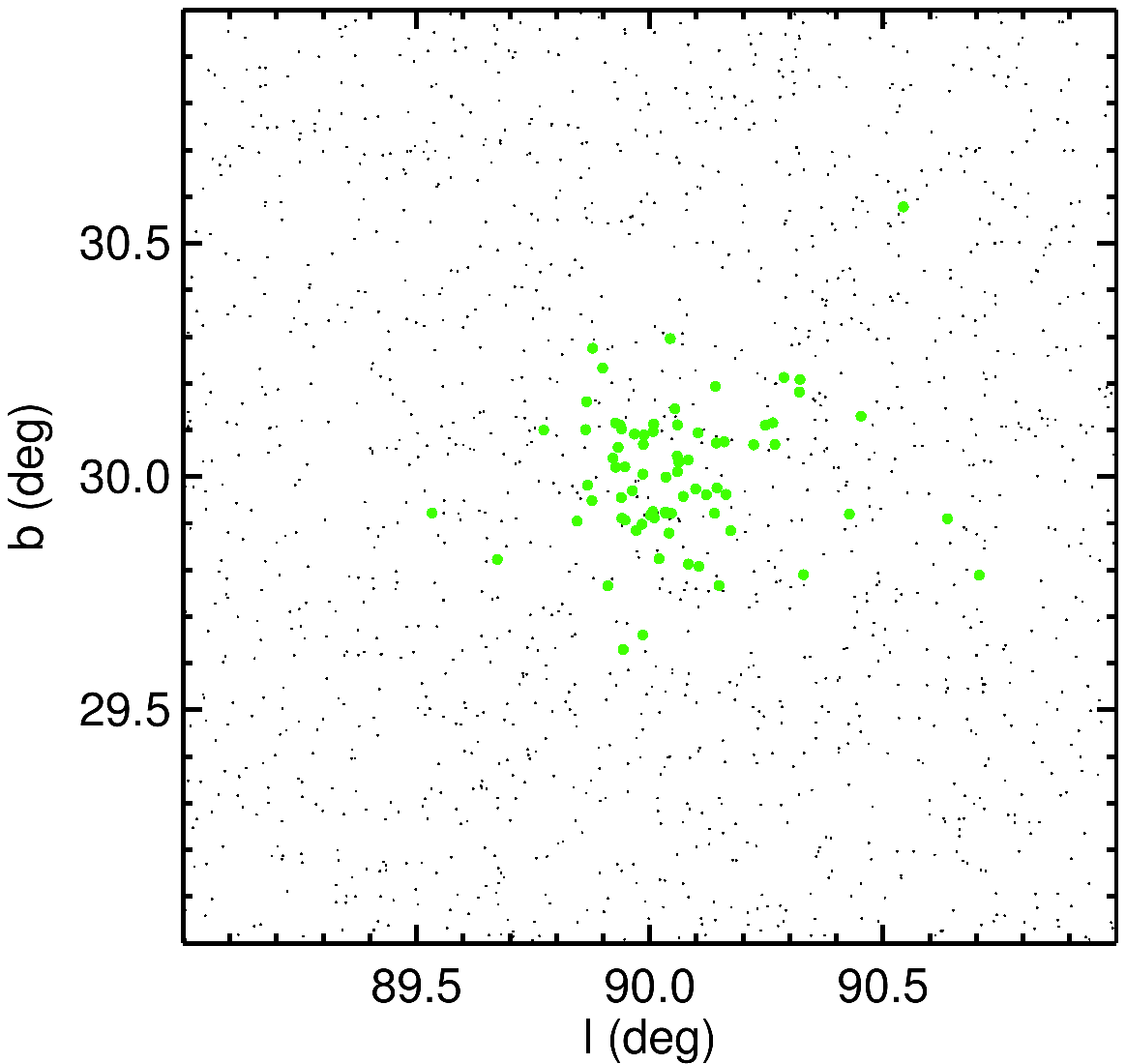} 
\vspace{0.2cm}

\includegraphics[width=0.4\textwidth]{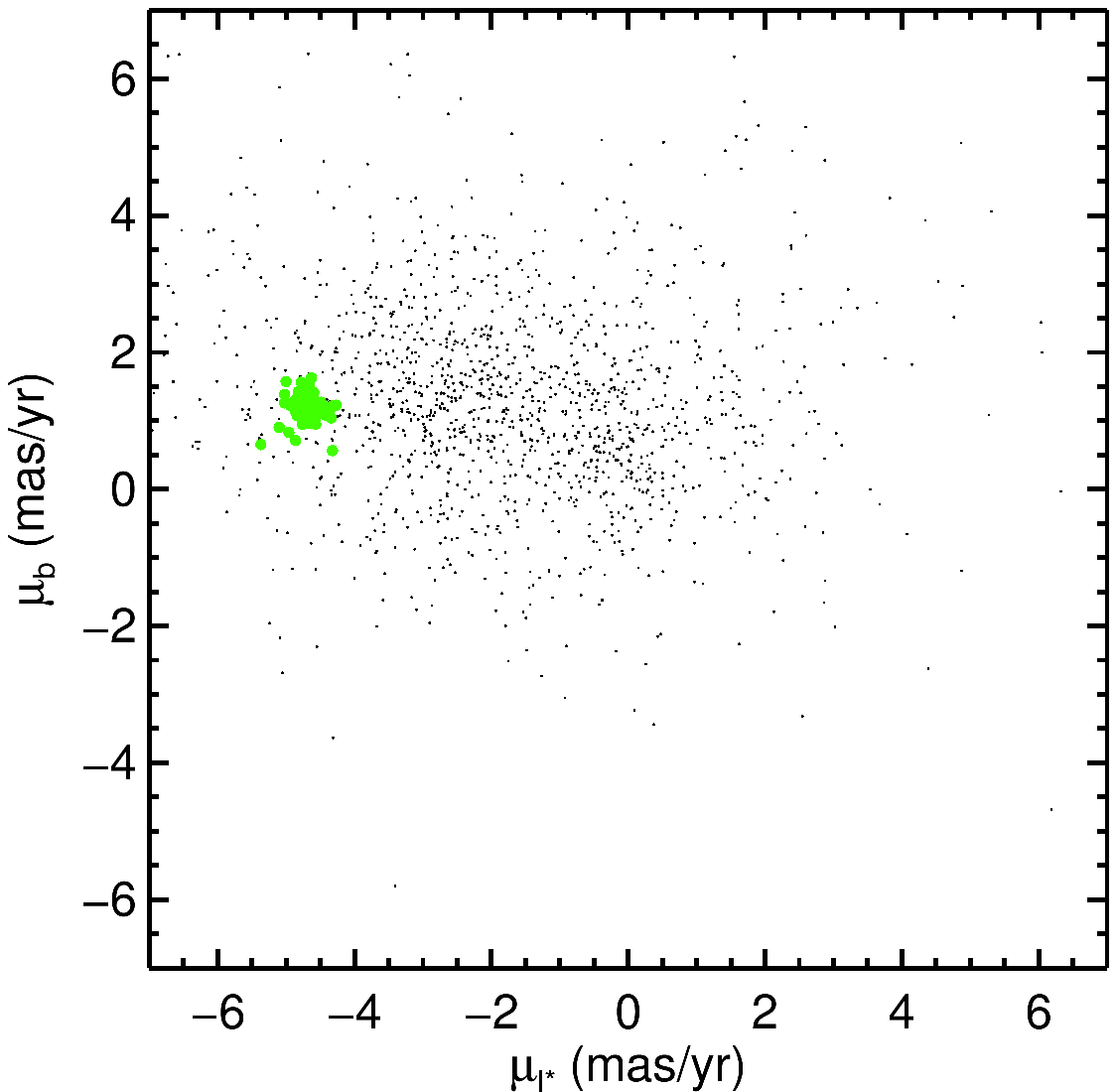} 
\caption{Sky (\emph{top}) and proper motion plane (\emph{bottom}) for the field of our fiducial \uf~in Table~\ref{t:example}. The stars belonging to the system are shown as green dots while the  background stars are in black.}
\label{f_points}
\end{figure}

The starting point of our detection procedure (Section~\ref{sec_Tools}) is the \ufs~projections in the sky and proper motion planes, thus it is essential to understand the behaviour of these projections of the systems and the background.  

Fig.~\ref{f_points} shows the stars in the customary $2\deg \times 2\deg$ field of view of our fiducial simulated system in the sky (top) and proper motion planes (bottom). The parameters of this system are listed in Table \ref{t:example}.  The stars belonging to the \uf~are coloured in green, while the background stars are in black. This system is hardly seen in the sky plane because it is very diffuse. But note how it is much more compact in the other plane. The compactness of the \ufs~in the proper motion plane is a general characteristic of most of our simulated \ufs~that improves considerably our search (Section~\ref{sec_Results}), being a fundamental advantage of the \gaia data.

Note also the very different nature of the background in these planes. In the sky the background is roughly constant, but becomes markedly non--uniform in the proper motion plane. This requires a special treatment when assigning significance to the peaks (Section~\ref{subsec_Probability}).

The apparent sizes of an \uf~in the sky and proper motion planes are set by its intrinsic size and velocity dispersion, combined with its distance from the Sun. These sizes can span a wide range in both planes. The half-light angular size is given by
\begin{equation}\label{e_lbsize}
\theta(\deg)\sim 0.0573\frac{r_{\rm h}(\pc)}{D(\kpc)}
\end{equation}
In our synthetic library, $r_{\rm h}$ varies between (see Table~\ref{t:example}) $5$ and $4\,000\pc$ and $D$ between 10 and $250\kpc$ (this is the approximate distance limit to detect at least $\sim$10 stars with {\em Gaia}, for a luminous \uf~with $L_V\sim 7\times10^4\Lsun$). This implies a range of apparent angular sizes of $[4"$, $23\deg$]. %

In the proper motion plane, the apparent size is
\begin{equation}\label{e_pmsize}
\Delta\mu(\masyr)\sim 0.211 \frac{\sigma_V(\kms)}{D(\kpc)}
\end{equation}
For values of $\sigma_V$ in the range $[1,500]\kms$ and again $D$ between 10 and $250\kpc$, we end up with a range of apparent sizes of $[0.0008,11]\masyr$ (but see below).

The right panel of Fig.~\ref{f_dmuth_NobsD} shows the sizes spanned in the sky and proper motion planes by the \ufs~in our library. As we will see in Section~\ref{sec_Tools}, these two parameters are the most important, together with the number of visible stars in the \uf, in determining the detectability of the system. We can see here that our library extends well beyond the spread covered by real systems. Note again that the boundaries of the library are sharp in this panel, as it is generated with \emph{apparent} parameters drawn at random from a uniform distribution in a logarithmic scale.

\begin{figure}  
\centering 
\includegraphics[width=0.42\textwidth]{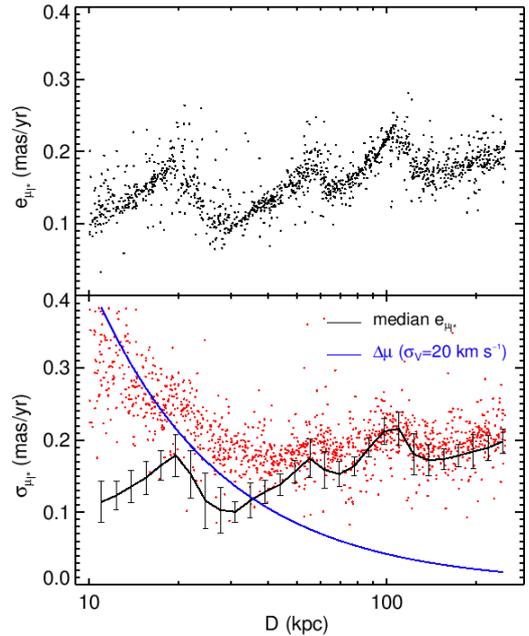} 

\caption{\emph{Top:} median error $e_{\mu_{\ell*}}$ in $\mu_{\ell*}$ for $\sim1300$ synthetic \uf~with velocity dispersions around $20\kms$ ($15<\sigma_V<25 \kms$) at different heliocentric distance. \emph{Bottom:} dispersion $\sigma_{\mu_{\ell*}}$ in the $\mu_{\ell*}$ proper motion of the same set of synthetic \uf~(red dots). The black curve shows the median error in $\mu_{\ell*}$ calculated in logarithmic bins from the top panel. The error bars correspond to the standard deviation. The blue line is the expected size according to Eq. \ref{e_pmsize} for $\sigma_V=20\kms$. }\label{f_errorsize20}
\end{figure}

It is also important to note that the apparent size of an \uf~in the proper motion space is greatly influenced by the observational errors. 
To illustrate this, we use a set of $\sim1,300$ simulated \ufs~located at different distances and with velocity dispersions between $15$ and $25\kms$. Each black dot in the top panel of Fig.~\ref{f_errorsize20} shows the error in $\mu_{\ell*}$ (similar for $\mu_b$) in each simulated \uf~computed as the median of all the individual stars errors in each \uf. The proper motion error slightly increases with distance, as one would naively expect due to the fainter magnitudes. But the error also oscillates with distance. 
This is because the \gaia performances depend on the magnitude and colour of the star and the type of stars in the \uf~that \gaia can detect, and thus the fraction of stars with certain magnitudes and colours, changes as it is observed at different distances (as seen in Section~\ref{subsec_GaiaMod}). One can see that the error has minima around 30, 70 and $135\kpc$. These are distances slightly larger than the ones at which a sudden increase in the number of stars of certain types occurs. They correspond to the main sequence turn-off, the extreme horizontal branch and the horizontal branch, as discussed previously (gray shaded stripes in Fig.~\ref{f_cmd_bassic}). 
%This is because the type of stars of the \uf~population that \gaia can detect changes as it is observed at different distances (as seen in Section~\ref{subsec_GaiaMod}), meaning that the relative fraction of stars with certain magnitudes and colours changes (see Fig.~\ref{f_cmd_bassic}). As the \gaia performances depend on the magnitude and colour of the star, this translates in these observed oscillations. One can see that the error has minima around 30, 70 and $135\kpc$. These are distances slightly larger than the ones at which a sudden increase in the number of stars of certain types occurs (Fig.~\ref{f_Nobs}). They correspond to the main sequence turn-off, the extreme horizontal branch and the horizontal branch, as discussed previously (gray shaded stripes in Fig.~\ref{f_cmd_bassic}). 

The bottom panel of Fig.~\ref{f_errorsize20} shows the real size in the proper motion plane (red dots) computed as the standard deviation of the proper motion coordinate $\mu_{\ell*}$ ($\sigma_\mu\equiv\sigma_{\mu_{\ell*}}$) of the stars in each \uf. We also overplot the error in $\mu_{\ell*}$ ($e_{\mu_{\ell*}}$) at each distance (black curve) taken as the median error in logarithmic bins from the top panel. The blue line in this plot shows the expected size according to Eq.~\ref{e_pmsize} for a velocity dispersion of $20\kms$. We see that the sizes of the \ufs~decrease up to $\sim40\kpc$ and for larger distances they follow the oscillations due to the \gaia errors. Once the size of the \ufs~is dominated by the observational error, the apparent size oscillates between 0.1 and $0.2\masyr$. For smaller velocity dispersions, e.g. $\sim 5\kms$, the errors dominate already at a distance of $10\kpc$. Therefore, the minimum apparent size of the \ufs~is set by the observational errors. For the range of parameters explored here, this is above $0.1\masyr$ in $99.5$ per cent of the cases. In what follows we take $\sigma_\mu$ instead of $\Delta\mu$ as a better measure of the apparent size of the \uf~in proper motion space.

As seen in Section~\ref{subsec_GaiaMod}, the errors in the angular coordinates in the sky are around 0.05-0.4 $\rm{mas}$, which is negligible compared to the apparent sizes of the \ufs. For this reason, we do not observe a similar effect in the sky plane.

%------------------------------------------------------------------------------------------------------------------

\section{The detection tools}
\label{sec_Tools}

In this section, we present all the different elements that compose our detection method to identify \uf~candidates against the background.

Our strategy is as follows. We consider fields of view of $2\deg\times 2\deg$ in the sky. We first detect over-densities independently in the sky and the corresponding proper motion planes. For this we use the Wavelet Transform (WT, see Section~\ref{subsec_Wavelet}). We do this for over-densities of different amplitudes and sizes,
and keep the most significant ones.
After this, we perform what we call the cross-match of peaks (Section~\ref{subsec_CrossCorr}). This consists in counting how many stars belong {\em simultaneously} to a certain peak in the sky and to a certain peak in proper motion space. We do this for {\em all pairs} of peaks of any size between both spaces. For each cross-match, we finally compute the probability that the observed number of common stars is just a coincidence (Section~\ref{subsec_Probability} to \ref{sec_independent}). Cross-matches with a low probability are selected as possible \uf~candidates. Below we detail each of the steps of our method.

One may wonder why this separate treatment for the sky and proper motion planes. After all, what we are looking for is a single peak in the 4-D space of positions in the sky and proper motion planes. This is because of the very different nature of these two planes, which results in the impossibility of having a natural metric in the combined space. Any metric will imply the introduction of an arbitrary dimensional scale which will limit the nature of the systems found. This is why we have preferred to work on the sky and proper motion planes separately, and then use the cross match procedure to relate peaks. The peaks that we do identify correspond to single peaks in the combined 4-D space, but not necessarily using a unique metric, as our combination of different wavelet scales in both planes allows for a larger range of identified peaks than if using a single metric.

Although the whole detection process might seem complex, it is quite straighforward from the computational point of view. The entire algorithm takes a total of 40 s to run for our fiducial field of $2\deg\times2\deg$ in a single Intel(R) Core(TM) i7-3770 CPU @3.40GHz. This might change depending on the LOS but, as a first approximation, the celestial sphere above $b=30\deg$ would require 86 h of CPU time, which in fact can be spread into several CPU for different LOS.

%--------------------
\subsection{Wavelet analysis}
\label{subsec_Wavelet}

\begin{figure*}  
\centering 
\includegraphics[width=1.\textwidth]{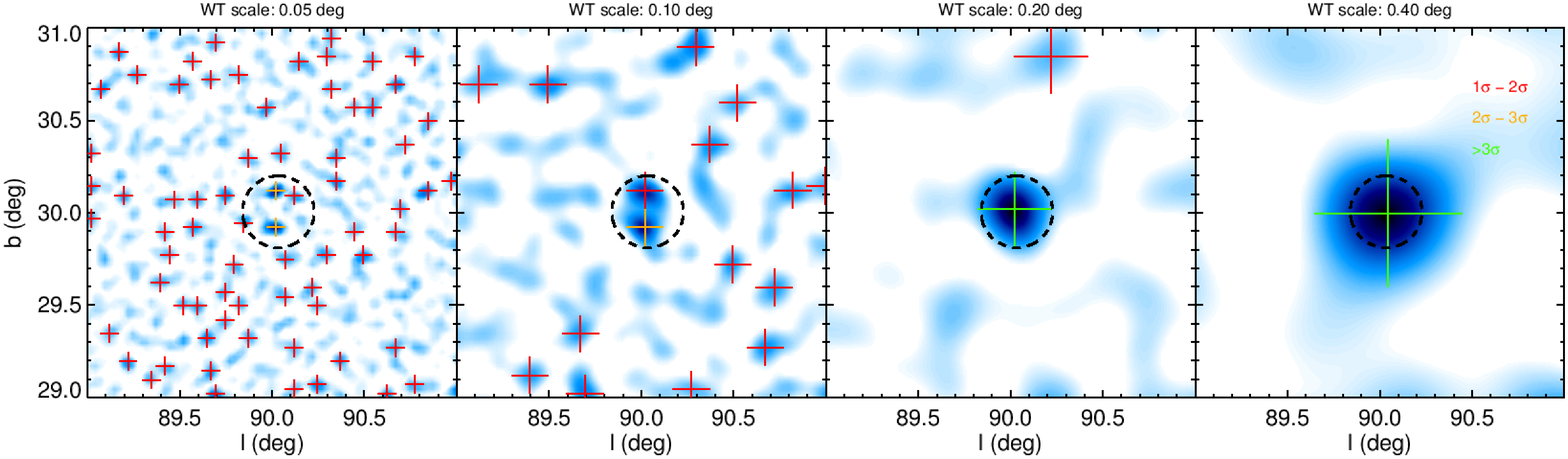}
\vspace{0.02cm}

\includegraphics[width=1.\textwidth]{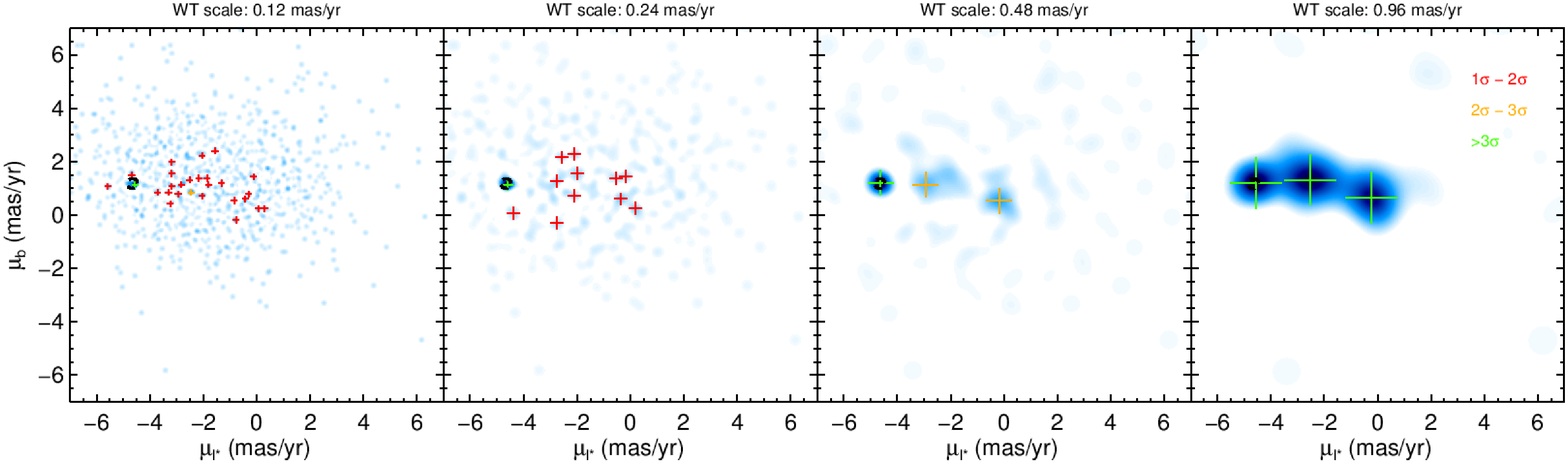} 
\caption{Wavelet Transform (WT) at different scales for our fiducial \uf~for the sky (\emph{top}) and proper motion planes (\emph{bottom}). The black dashed circle shows the true position and size of the system in each plane. The circle in the proper motion plane is very small but can be seen better in the smaller scales. The position is calculated as the median of the coordinates (positions and proper motions) of the stars in the \uf~that are observed by {\em Gaia}, while the size of the circle is taken as the maximum between the standard deviation of the coordinates. Red, orange and green crosses indicate peaks at between 1 and 2, between 2 and 3, and $>3$ $\sigma$ significance, respectively.}\label{WT}
\end{figure*}

To detect over-densities in the sky and proper motion planes like the ones of Fig.~\ref{f_points}, we use the WT \citep{Starck02}. This can be thought of as a ``localized'' Fourier transform that gives information about certain frequencies and where in the image these frequencies are located. Due to the wide range of apparent sizes of our simulated \ufs~(see Section~\ref{subsec_planes}), our method needs to be able to detect over-densities of different sizes. In the application here a discrete set of frequencies (i.e. scales) are probed and we get information about the localization of those particular structures. We use here the {\em \`a trous}  (``with holes'') variant of the WT \citep{Starck02} which computes a discrete set of scale-related ``views'' of a 2-D function or image. We have previously used this technique to detect moving groups in the stellar velocity distribution of the Solar Neighborhood and surroundings \citep{Antoja08,Antoja12}. To perform the calculations we use the MR software developed by CEA (Saclay, France) and Nice Observatory.

Although the WT works at a specific scale, it can identify over-densities within some range in size. Nevertheless, it is important to realize that we are probing a discrete set of scales in the images, and therefore, it is critical to choose those scales wisely. We explore 4 logarithmically spaced scales in each plane within the ranges found in Section~\ref{subsec_planes}. For the sky, as we are dealing with fields of $2\deg\times2\deg$, we have chosen the scales $0\fdg05$, $0\fdg1$, $0\fdg2$ and $0\fdg4$. Even though the higher scale puts a limit on the maximum size of an \uf~that can be detected in principle, the innermost parts of the more luminous \ufs~can still be detected, even if they have larger angular sizes. For the proper motion plane, we use scales of $0.12$, $0.24$, $0.48$ and $0.96$ $\masyr$. Here, what we are missing are exceptional cases with extremely high velocity dispersion which are very close.

An example of the WT planes in the sky for our fiducial \uf~in Table~\ref{t:example} is shown at the top part of Fig.~\ref{WT}, while the bottom part shows the proper motion plane. In each case, the four scales mentioned are shown. The blue colours are proportional to the values of the WT. 

After the WT, we search for relative maxima to detect the over-densities. The algorithm computes the Wavelet Probability (WP), that is the probability that the detected over-densities in the wavelet space are not due to Poisson noise. For this it uses a model for this type of noise in wavelet space. This is done by first using the Anscombe transform \citep{Anscombe48} that converts a signal with Poisson noise into Gaussian noise, for which the treatment in the WT planes is more straightforward (see \citealt{Starck02} and references therein). 
Here we will consider only WT peaks that have a WP of being real detections of $WP\ge 99.7$ per cent (green crosses in Fig.~\ref{WT}), $95.4\le WP< 99.7$ (orange crosses) and $68.2\le WP< 95.4$ (red crosses), similar to $>$\3s, {\2s}--{\3s} and {\1s}--{\2s} significance levels in the Gaussian case, respectively. The size of the crosses in this figure indicates the size or scale that is being probed in each WT plane (also indicated in the top of the plots). There is an additional condition on the over-densities: they should have at least 5 stars to be considered a peak. 

The \ufs~are optimally detected (i.e. with higher WP) when the scales probed are similar to their apparent sizes. In the example of Fig.~\ref{WT} a black circle indicates the position and extension of the \uf~in the sky plane and in the proper motion plane. In the sky images, for scales that are smaller than the apparent size of the \uf~(two first panels), some peaks are detected inside the region occupied by the \uf~but with low WP (\1s or \2s, red and orange crosses). For larger scales, which in this case are similar to the apparent size of the \uf, the detection is above \3s (green crosses). In other cases, the detection is always below \3s or below \2s because the \ufs~can be very diffuse in this plane, as already highlighted in Section~\ref{subsec_planes}. 

On the other hand, the \uf~is very compact in proper motion space, and it stands out as an over-density for all scales studied in the left part of the panels. In our example, the fiducial \uf~is always detected above \3s (green crosses). In other cases, the best detection is for a particular scale that is close to the apparent size of the  \uf.

Note also how in both planes, a number of low--WP random detections appear (most of red crosses in Fig.~\ref{WT}). Because of this, we need to discard false over-densities and keep only good \uf~candidates (Section~\ref{subsec_CrossCorr}). Besides, in the proper motion case, two over-densities with high WP are also detected in the center of the distribution for the two largest scales. These correspond to the peaks of the background distribution, which as we have seen in Fig.~\ref{f_points}, is not uniform. Note that the distribution of background stars in the proper motion plane will be different for each LOS and therefore, its centroid will shift to different positions in this plane. This does not occur for the sky plane which presents a uniform background.

%--------------------

\subsection{Cross-matching peaks in the two planes}
\label{subsec_CrossCorr}

So far, we have detected peaks in the sky and proper motion planes, separately. However, contrary to false detections, an \uf~is an over-density in the 4-D combined space $l$-$b$-$\mu_{\ell*}$-$\mu_b$. This is precisely the feature that we need to exploit to identify \ufs, beyond what has been currently achieved. 

To do this, we list the stars contributing to the peaks identified separately in the sky and proper motion planes. By stars belonging to a certain peak, we mean those that are enclosed in a circle around the peak with the radius of the WT scale in the considered plane. In practice, because \gaia is a point source catalogue, we can identify the stars by their {\em id} number. Then we see whether a large fraction of these stars belong {\em simultaneously} to a certain peak in the sky and a peak in the proper motion plane. We call this ``cross-match of peaks''.  This cross-match is done for every peak and at every scale in the sky, compared to every peak at every scale in the proper motion plane. 

The computation of the probability of having this cluster of common stars occurring by chance is computed as described in Section~\ref{subsec_Probability}. In Section~\ref{subsec_threshold} we explain how we filter out false detections. Because each \uf~can be detected in more than one scale, we also need to keep only independent detections. This is explained in Section~\ref{sec_independent}.

%--------------------

\subsection{Assessing the probability of the detections}
\label{subsec_Probability}

Here we describe the statistics machinery that we devised to assess the probability of detection, i.e. compute which detections have a very low probability of occurring by chance.

We are interested in $P(N_{\rm com} | \langle N_{\rm com}\rangle)$, i.e. the probability of observing a certain number of common stars $N_{\rm com}$ in a peak in the sky and a peak in the proper motion plane, given the expected number of common stars $\langle N_{\rm com}\rangle$. This probability\footnote{Do not confuse this probability for the combined sky and proper motion planes $P$, with the wavelet probability WP used in Section~\ref{subsec_Wavelet}} is simply given by the Poisson probability distribution function

\begin{equation}\label{eq_p}
P\equiv P(N_{\rm com} | \langle N_{\rm com}\rangle)=\mathrm{Poisson}(N_{\rm com} | \langle N_{\rm com}\rangle).
\end{equation}

An estimate of $\langle N_{\rm com}\rangle$ is given by: 

\begin{equation}\label{e:Ncom_expected}
\langle N_{\rm com}\rangle = \langle N_{\rm sky}\rangle \int_{A_\mu} \rho(\mu_{\ell*},\mu_b)\, d\mu_{\ell*} d\mu_b,
\end{equation}
where $\langle N_{\rm sky}\rangle$ is the expected number of stars in the $l-b$ peak and $\rho(\mu_{\ell*},\mu_b)$ describes the (normalized) number density of stars in the proper motion plane, both under the assumption that no \uf~is present. $A_\mu$ indicates the area of the peak over which we are integrating, which is a circle with a radius given by the WT scale, centered on the $(\mu_{\ell*},\mu_b)$ coordinates of the peak in question. For convenience, we use hereafter the logarithm of the probability, $\ln P$.

For simplicity, we assume that the background density in the $l-b$ plane is uniform, which is reasonable for the field size used, and therefore  $\langle N_{\rm sky}\rangle= N_{\rm BG} (\pi r_{\rm sky}^2)/A_{\rm T}$, where $A_{\rm T}$ is the total area of the field in the sky plane (in our case 4 \sqdeg), $r_{\rm sky}$ is the wavelet scale in the plane of the sky and $N_{\rm BG}$ is the number of background stars in the field. The latter is computed from the observed data itself, by taking the 8 fields adjacent to our problem field, with the same total area. For each of these fields we compute the total number of stars and we take the median. This is a better estimation of the number of background stars than the total number of stars in the considered field, specially in cases of luminous \ufs~that have a number of observed stars that is not negligible compared to the number of background stars.

In the proper motion plane, however, it is crucial to account for the fact that the density is not constant and this is achieved by the integral term in Eq.~\ref{e:Ncom_expected}, which, multiplied by $\langle N_{\rm sky}\rangle$ gives the number of common stars expected to lie within the area of the detected peak.

The distribution of stars in the proper motion plane $\rho(\mu_{\ell*},\mu_b)$ is different depending upon the direction on the sky, so it must be computed independently for each field. We do this from the observed data itself, by taking the mentioned 8 adjacent fields with the same total area. For each of these fields we compute the density as a (normalised) 2-D histogram in the $\mu_{\ell*}-\mu_b$ plane with a pixel size of $0.8\masyr$. Taking these eight 2-D histograms, we compute the pixel-by-pixel median density to obtain a statistically reliable estimate in each of the matrix cells, minimizing the effect of outliers.
With this, we are assuming that the distribution of background proper motions remains similar among these adjacent fields. This is indeed the case. For instance, variations in the median proper motion in longitude and latitude among the adjacent fields are in general smaller than the pixel size. 
We numerically evaluate the integral in Eq. \ref{e:Ncom_expected} using the trapezoid rule in 2-D and bi-linear interpolation on the median density matrix.

It could happen that one or various of the adjacent fields contains \ufs. This would yield a wrong estimation of $N_{\rm BG}$ and $\rho(\mu_{\ell*},\mu_b)$. The fact that we use the median of the 8 fields helps to alleviate this issue. However, in case very luminous \ufs~are present, our algorithm checks if the number of stars in one of the adjacent fields is significantly larger than in the others. This is done by checking that the dispersion in the number of stars in the 8 fields is not larger than 2.5 times the square root of the median. In this case, the algorithm could be re-run without the field in question.
 
Instead of using the probability of Eq.~\ref{eq_p}, we can also use the significance $s$, defined as the number of times above the expected value of the distribution, scaled to the dispersion of the distribution

\begin{equation}\label{eq_s}
s=\frac{N_{\rm com}-\langle N_{\rm com}\rangle}{\sqrt{\langle N_{\rm com}\rangle}}.
\end{equation}
The advantage of using $s$ instead of $\ln P$ is that $s$ is positive and it increases for more relevant detections.

%--------------------
\subsection{Setting a threshold probability for detection}
\label{subsec_threshold}

\begin{table*}
 \caption{Detections, real and false, as percentage of the number of \ufs~used in each LOS.}\label{t_false}
\centering
\begin{tabular}{cc|rr|rr|rr|rr|rr}\hline\hline
$\ln P_{\rm thres}=$&&\multicolumn{2}{|c|}{$-3.0$}&\multicolumn{2}{c|}{$-6.0$}&\multicolumn{2}{c|}{$-9.0$}&\multicolumn{2}{c|}{$-12.0$}&\multicolumn{2}{c}{$-15.0$}\\
  l  & b&$\%_{\rm rec}$&$\%_{\rm false}$&$\%_{\rm rec}$&$\%_{\rm false}$&$\%_{\rm rec}$&$\%_{\rm false}$&$\%_{\rm rec}$&$\%_{\rm false}$&$\%_{\rm rec}$&$\%_{\rm false}$\\\hline
 90  &   30  &   86.8  &  150.2  &   86.7  &  105.0  &   85.4  &   50.3  &   83.4  &    0.3  &   81.7  &    0.0  \\
 90  &   42  &   85.7  &   44.9  &   85.4  &   38.7  &   84.3  &    1.2  &   83.0  &    0.0  &   81.6  &    0.0  \\
 90  &   55  &   84.6  &   37.4  &   84.4  &   35.9  &   83.6  &    1.8  &   82.6  &    1.1  &   81.8  &    0.3  \\
 90  &   68  &   81.4  &   39.0  &   81.2  &   38.9  &   80.8  &   27.8  &   80.2  &    0.3  &   79.3  &    0.0  \\
 90  &   80  &   84.6  &    7.8  &   84.2  &    0.5  &   83.4  &    0.1  &   82.8  &    0.0  &   82.2  &    0.0  \\
170  &   30  &   87.7  &   68.6  &   87.4  &   67.5  &   85.9  &    0.4  &   84.6  &    0.1  &   83.4  &    0.0  \\
170  &   42  &   84.7  &    6.3  &   84.4  &    0.6  &   83.8  &    0.2  &   83.2  &    0.0  &   82.6  &    0.0  \\
170  &   55  &   79.9  &   47.6  &   79.7  &   43.8  &   79.3  &    0.4  &   78.8  &    0.0  &   77.8  &    0.0  \\
170  &   68  &   82.2  &    1.7  &   82.1  &    0.5  &   81.7  &    0.0  &   81.5  &    0.0  &   80.6  &    0.0  \\
170  &   80  &   82.9  &    3.2  &   82.6  &    0.4  &   81.5  &    0.0  &   81.1  &    0.0  &   80.8  &    0.0  \\
\hline
\end{tabular}
\end{table*}

We are only interested in those detections which have very low probability (very negative $\ln P$) \emph{and} also a number of common stars\footnote{The last condition is required in order to select only over-densities but not under-densities.} $N_{\rm com}>\langle N_{\rm com}\rangle$. However, the central peak(s) of the background in the proper motion plane can appear also as a detection. This is because we estimate its expected number of stars using the adjacent fields and any small fluctuation above this value can give a significant detection, though with a larger $\ln P$. To filter peaks that correspond to the background and not to the \ufs, we can conservatively select a relatively low (very negative) threshold value for $\ln P$, below which we consider detections to be relevant. However, we must realize that, as we lower the threshold value $\ln P_{\rm thres}$, although we eliminate spurious peaks, we start losing relevant detections, so this is a compromise between false positives and losing bona fide peaks.

We have explored this compromise on various LOS's. In Table~\ref{t_false} we list the percentage of recovered \ufs~$\%_{\rm rec}$ and of false detections $\%_{\rm false}$ from the total number of tested \ufs~, as a function of 5 different values for the threshold and 10 different LOS's. For values above -9.0 there are several fields where the percentage of false detections is above 20 per cent. For a threshold of -12.0, all false detections are at most 1.1 per cent. Although we could choose a $\ln P_{\rm thres}$ in between those two to make it the most optimal, we conservatively choose $\ln P_{\rm thres} = -12$.

%--------------------
\subsection{Independent Detections}\label{sec_independent}

As explained before, the cross-match of peaks is done for all peaks at all scales and a given \uf~can be detected in more than one scale. This means that we need to select which of the many detections made in a given field, are in fact independent detections, i.e. different objects. 

We first organize all detections by increasing $\ln P$, choose the detection with the lowest $\ln P$ and compare its $l-b$  and $\mu_{\ell*}-\mu_b$ coordinates with the remaining detections. Now, we choose the next independent detection as the one with the lowest $\ln P$ that lies, both in the sky and proper motion planes, at a distance \emph{larger} than the sum of the WT scales of the two detections (i.e. they do not overlap). We repeat this procedure until we have gone through all the available (relevant) detections. 

\begin{figure}
\centering
\includegraphics[width=0.4\textwidth]{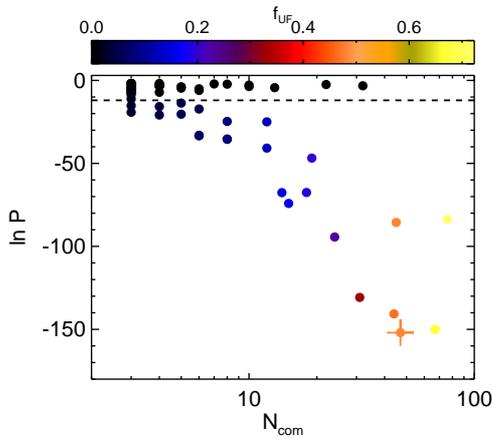}
\caption{$\ln P$ as a function of the number of common stars $N_{\rm com}$ for our fiducial field and \uf. The color scale is proportional to $f_{\rm UF}$, i.e. the fraction of recovered stars from each \uf. The dashed horizontal line mark the line below which we consider detections as relevant. 
The detection with the lowest $\ln P$ (marked with a cross) is what we take as the best \emph{independent} detection (see text for details).
}
\label{f_test0}
\end{figure}

To illustrate the behaviour of $\ln P$ and the selection of independent detections, in Fig.~\ref{f_test0}, we plot for our fiducial \uf, $\ln P$ as a function of the number of common stars $N_{\rm com}$ for all detections in this field, that is the results of cross-matching all peaks at all scales in the sky and proper motion plane. 
In the plot we use a color scale proportional to $f_{\rm UF}$, defined as the number of stars from those $N_{\rm com}$ that truly belong to the \uf~divided by the total number of stars originally in the \uf. In other words, $f_{\rm UF}$ is the fraction of recovered stars from each \uf. Dots below the horizontal dashed line are relevant detections, i.e. with $\ln P$ below $\ln P_{\rm thres}$ (Section~\ref{subsec_threshold}).

There is a correlation of $\ln P$ with $N_{\rm com}$. As expected, detections with larger numbers of common stars have on average lower $\ln P$. There is also a sequence that moves across the plot above the threshold. This corresponds to peaks in the proper motion background (note that they are black points, i.e. with no stars belonging to the \uf). As explained before, these detections are filtered by our threshold.

Also, detections with the largest values of $f_{\rm UF}$ have low $\ln P$, i.e. they are significant detections. However, as the number of stars in common increases, the value of $\ln P$  decreases, reaches a minimum and then increases again. The minimum value occurs for detections at the optimum scales in the sky and proper motion planes. It is in this case that a large fraction of \uf~stars lie in the detected peak inside the WT scale, and the background is sufficiently low, so that the difference between the \emph{observed} and expected number of common stars in the peak is maximal.
Increasing the WT scale past the optimum values causes the inclusion of more stars of the \uf~in the peak but also more stars of the background that might not necessarily belong to both peaks in the sky and proper motion plane  simultaneously, and hence, this causes $\ln P$ to go back to larger values. This is a very convenient behaviour which allows us to select detections at the optimum WT scales. 
The detection with the lowest $\ln P$ (marked with a cross) is what we take as the best \emph{(and in this case, only)} \emph{independent} detection.
Finally, notice that in this example no false positives are picked up. 

%----------------------------------------------------------------------------------------------------------------

\section{Results} 
\label{sec_Results}

As we have seen in Section~\ref{subsec_library}, we face a 9-D parameter space. Even with our library of more than 30\,000 different synthetic \ufs, it is clear that we can cover only a limited amount of this vast hypervolume.

To explore this space with some order, we will rely first on a series of carefully curated ensembles of cases. In each one, all parameters, except two, will be kept fixed (Section~\ref{subsec_PhysParam}). This allows us to take 2-D sections of the original parameter space. Then we will identify in Section~\ref{subsec_RedParam}, a reduced number of combinations of the original parameters that our detection procedure depends on directly, and which we call ``efective parameters''. In Section~\ref{subsec_limits}, we explore the limits and completeness of our method in the space of effective parameters, as well as in some of the most relevant original parameters. The effect introduced by changing the background level as we look at different LOS's is discussed in Section~\ref{subsec_bg}.

%--------------------

\subsection{The physical parameter space}
\label{subsec_PhysParam}

\begin{figure}
\centering
\includegraphics[width=0.38\textwidth]{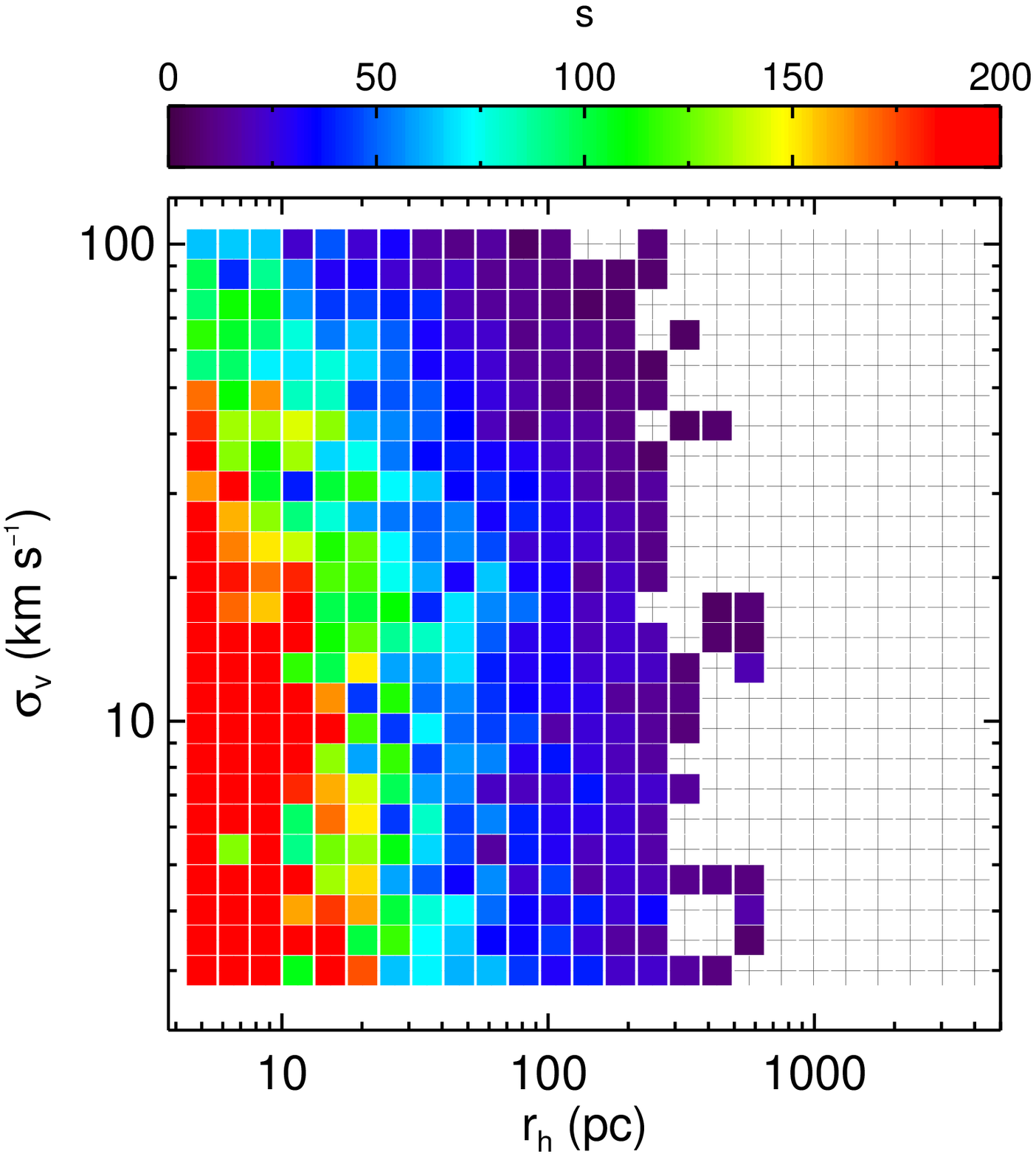} 
\vspace{0.45cm}

\includegraphics[width=0.38\textwidth]{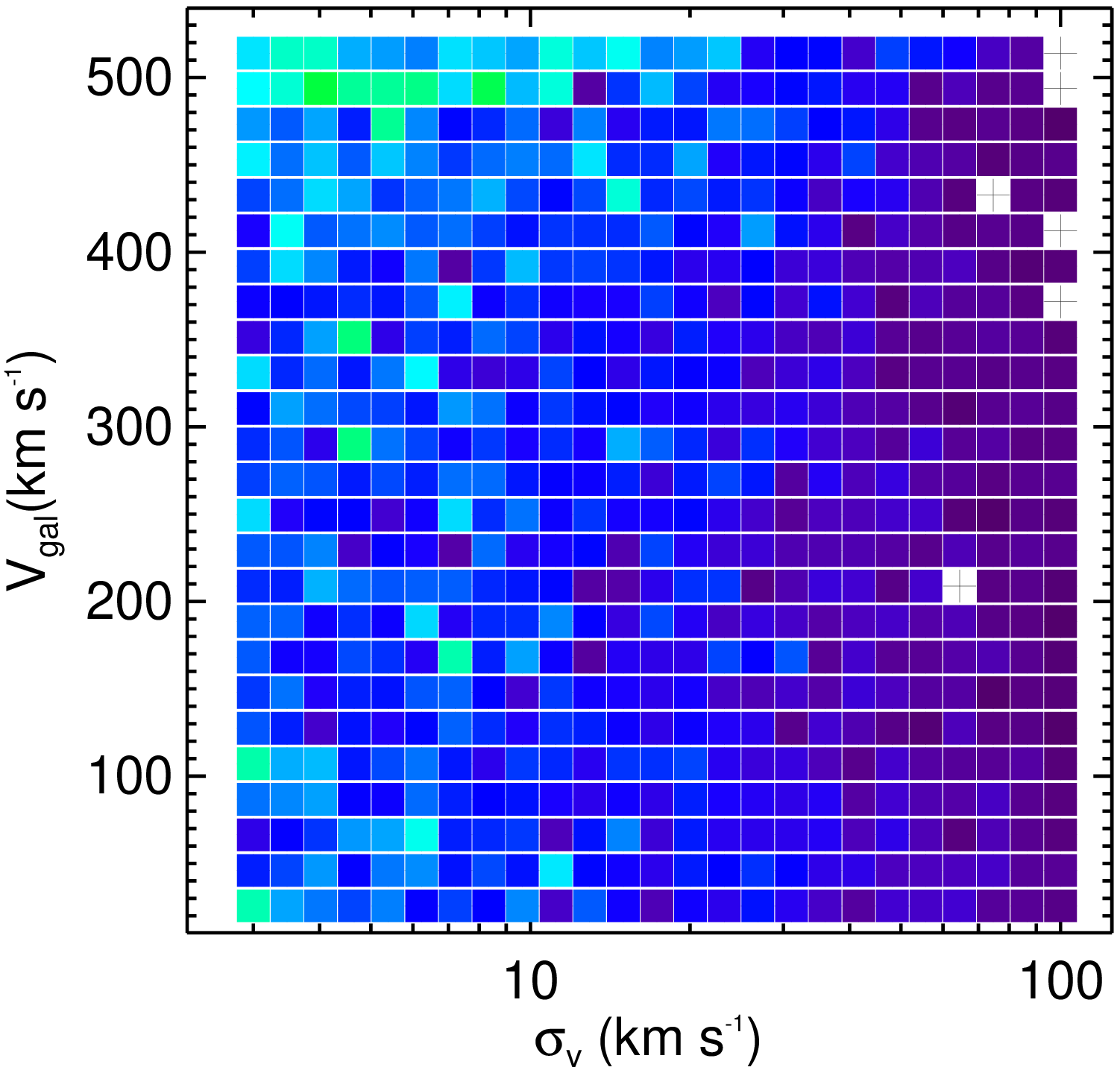}
\vspace{0.45cm}

\includegraphics[width=0.38\textwidth]{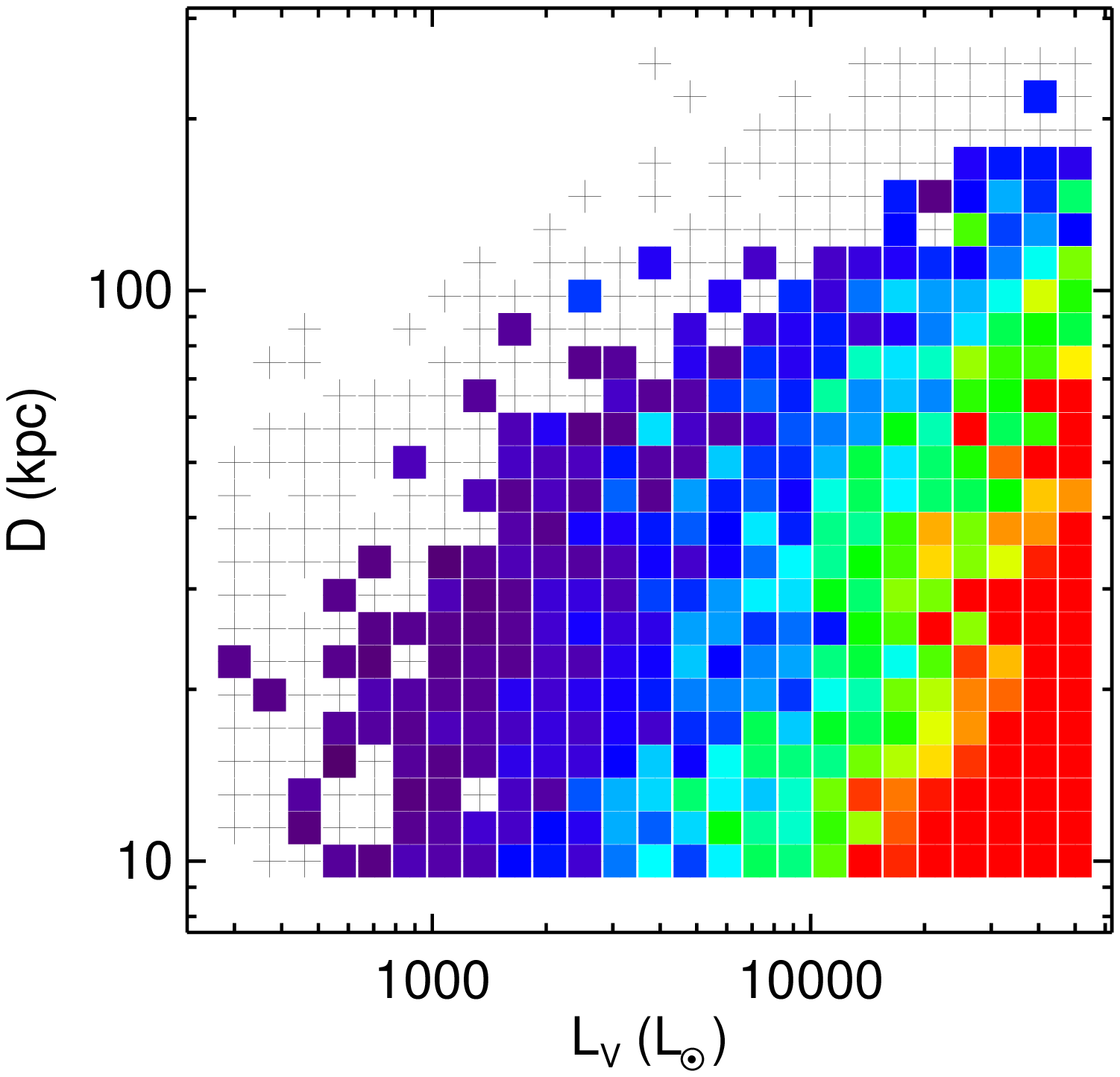}

\caption{Detectability tests run with several ensembles of \ufs~with only two varying parameters: $r_{\rm h}$ vs $\sigma_V ($\emph{top}), $\sigma_V$ vs $V_{\rm gal}$ (\emph{middle}), $L_V$ vs distance D (\emph{bottom}). The color scale indicates the detection significance $s$. \ufs~with significance over 200 have been plotted with a colour saturated at this value. Black crosses indicate \ufs~that were not detected.}
\label{f_tests}
\end{figure}

\begin{table*}
 \caption{Values for the fixed parameters in the ensembles of \ufs~shown in Fig.~\ref{f_tests}. }\label{t_fixed}
\centering
\begin{tabular}{ccrcccccc}\hline\hline
          $L_V$         &$r_{\rm h}$& $\sigma_V$ & $D$ & $l$    & $b$   &$V_{\rm gal}$ &$\phi_V$ &$\theta_V$\\
        ($\Lsun$)       &($\pc$) &($\kms$) &($\kpc$)&($\deg$)&($\deg$)&($\kms$)&($\deg$)&($\deg$) \\\hline
 $5\times10^3$   &  80   &  5    & 20&    90&    30&      300.&  0.  &0.         \\
\hline
\end{tabular}
\end{table*}

In the ensembles of tests presented here, we vary only 2 parameters, keeping the other 7 parameters constant. The values for the fixed parameters are shown in Table \ref{t_fixed}.

%-------------------
\subsubsection{The $r_{\rm h}$ vs $\sigma_V$ plane}
\label{subsubsec_rhsigmav}

In the first test we use $625$ synthetic \ufs~with varying $r_{\rm h}$ and $\sigma_V$. As indicated in Table \ref{t_fixed}, all \ufs~are located at a fixed position in the sky at $l=90^\circ$, $b=30^\circ$, at a heliocentric distance of $20 \kpc$ and have a luminosity of $L_V=5\times10^3\Lsun$. These are approximately the mean values for the observed \ufs. The velocity dispersion varies logarithmically between 3 and $100\kms$ and the half-light radius between $5\pc$ and $4\kpc$. Due to stochastic variations in each realization, despite the luminosity and distance being constant, the number of detectable stars $N_{\rm obs}$ varies between 38 and 207. 
The results of the test are shown in the top panel of Fig.~\ref{f_tests}. Each symbol (squares and crosses) in this plot corresponds to one simulated \uf. The color scale in the panels is proportional to the detection significance (Eq.~\ref{eq_s}). Black crosses indicate \ufs~that were not detected. From this plot we can evaluate the detection limits as a function of $r_{\rm h}$  and $\sigma_V$.

\ufs~with $r_{\rm h}$ larger than $\sim 600\pc$ are not detected (for this fixed distance and luminosity). This is because their apparent size in the sky is very big, making them extremely diffuse. In fact, $r_{\rm h} = 700\pc$ results in an angular size equal to the sky fields that we are using for our analysis ($2\deg\times2\deg$). We also notice that for velocity dispersions below $20\kms$ the detection significance depends mainly on the half-light radius (vertical contours). This is because in this regime the apparent size of the \uf~in proper motion is in fact constant and set by the observational errors (Section~\ref{subsec_planes}). 
Above this velocity dispersion, the contours bend slightly to the left, meaning that for a given size in the sky, the detection is more significant for lower velocity dispersions. Note that we are exploring velocity dispersions up to $\sim100\kms$, i.e. significantly larger than the typical velocity dispersion of $\sigma_V\sim5\kms$ of known \uf~and classical dSph galaxies \citep{McConnachie2012}.

%-------------------

\subsubsection{The $\sigma_V$ vs $V_{\rm gal}$ plane}
\label{subsubsec_sigmavVgal}

The middle panel of Fig.~\ref{f_tests} is a test with 625 \ufs, where the velocity dispersion $\sigma_V$ and the modulus of the velocity vector $V_{\rm gal}$ are varied. In this case, $V_{\rm gal}$ is varied linearly instead of logarithmically. Notice that in this ensemble, we only change the position and spread of the \uf~peak in the proper motion plane. In particular, we have chosen the values for the fixed velocity angles ($\theta_V$ and $\phi_V$), so that the \uf~peak position moves horizontally across the proper motion plane as we vary $V_{\rm gal}$, covering all possible contrasts between background and UFDG, and coinciding with the background peak for $V_{\rm gal}\sim~150\kms$. 

Note how for a fixed value of $V_{\rm gal}$, the best detections are the ones for lower velocity dispersions, which produce more concentrated peaks. Besides, something that immediately stands out from this plote, compared with the others shown in Fig.~\ref{f_tests}, is the shallow variation in the detection significance across the entire part of this plane. At the horizontal region around $V_{\rm gal}\sim 150\kms$, we see that the significance is the lowest, as we expected, but this is a very subtle effect. This lack of sensitivity indicates that, although they play a role, these two parameters (and specially $V_{\rm gal}$) have little effect on the detectability of the \uf.

%-------------------
\subsubsection{The $L_V$ vs $D$ plane}
\label{subsubsec_LD}

The bottom panel of Fig.~\ref{f_tests} shows the results of the experiment where luminosity and distance were varied between $3\times10^2$ and $5\times10^4\Lsun$, and between 10 and $250\kpc$, respectively. There are 509 \ufs~in this test. Their mass--to--light ratio $M/L$ is between $~20$ and $4~\times10^3$. Their observable number of stars $N_{\rm obs}$ varies between 2 and $~6\,000$. Note that here we consider a lower value for the minimum $N_{\rm obs}$ than 10 as indicated in Table~\ref{t:example} to sample in detail the detection limit. We find, however, that the minimum number of $N_{\rm obs}$ that gives a positive detection is 5 for this particular example.

In this test, there are several competing effects. For a fixed luminosity, as we increase the distance, the size in the sky and proper motion planes decreases, which favours identification, but on the other hand, the number of visible stars also decreases, which makes identification harder. The first effect scales as $\propto 1/D$, while the second, being an individual star luminosity problem, scales as $\propto 1/D^2$. So, at large distances the latter dominates and we lose the \ufs, as seen in this panel. For instance, \ufs~with luminosities around $10^4\Lsun$ are not detected beyond $\sim100\kpc$. Also, given a fixed distance, more luminous objects are detected with higher significance. 

There is an interesting modulation in the colour contours in this panel. There are three leftward indentations of better detections at around 10, 25 and $60\kpc$, which are better seen in the red and light-blue colours. These features are not statistical fluctuations, but the result of the effect of the \gaia observational errors in measured proper motions. As explained in Sect~\ref{subsec_planes} (Fig.~\ref{f_errorsize20}), the size of the \ufs~in proper motion plane changes with distance in a peculiar way, presenting several minima at around the mentioned distances. At these distances, therefore, the \ufs~are slightly more concentrated in proper motion and, hence, easier to detect. The upper left part of the panel, which does not contain any coloured squares or black crosses, is the region where systems do not have stars that can be observed by {\em Gaia}.

From this simple tests, one can see that some properties of the \ufs~are more relevant for the detections. In particular, the luminosity and distance, which set the number of observable stars together with the apparent size in the sky, seem to have a larger impact on the significance of our detections, than the size in proper motion space and the position of the peak with respect to the background.

%-------------------------------------
\subsection{The ``effective'' parameter space}
\label{subsec_RedParam}

If we look at the essence of our problem devoid of its astronomical context, our task is to identify common peaks in two different planes, subject to a noisy and not necessarily uniform background. Seen as such, the key parameters upon which a successful detection depends are: the height of the peaks compared with the background level and the spread of the peaks, that is the apparent sizes of the \ufs~in the sky $\theta$ and proper motion planes $\sigma_\mu$, and the number of observed stars that they are composed of $N_{\rm obs}$, with respect to the background. The probability will also depend on the projection of the center of mass velocity in the proper motion plane, since this determines how close the \uf~peak appears to the center of the proper motions distribution, where the majority of the background contaminants lie.
   
One can also see that for a certain \uf~that has been detected in the optimal scales (that is almost all stars in the \uf~are enclosed inside the joint peak detection), $N_{\rm com}\sim N_{\rm obs}+N_{\rm BG,in}$, where $N_{\rm BG,in}$ is the number of background stars that fall inside the joint peak detection. Assuming that the number of background stars inside the joint peak is similar to the expected one, that is $N_{\rm BG,in}\sim\langle N_{\rm com}\rangle$, the significance of Eq.~\ref{eq_s} is equivalent to

\begin{equation}\label{eq_sr}
s\sim\frac{N_{\rm obs}}{\sqrt{ N_{\rm BG,in}}}.
\end{equation}

  As $N_{\rm BG,in}\propto \sim N_{\rm BG}\theta^2\sigma_\mu^2$, it follows from Eq.~\ref{eq_sr} that \ufs~that have
 the same ratio $N_{\rm obs}/\theta$ and all the rest of the parameters equal (including the LOS), have also approximately the same significance\footnote{Note also that \ufs~with the same $N_{\rm obs}/\theta^2$ or $N_{\rm obs}/\sigma_\mu^2$ (apparent ``surface'' density) do not have the same significance.}. The same applies to \ufs~with the same ratio $N_{\rm obs}/\sigma_\mu$. For this reason, given a certain LOS (we deal with different LOS in Section~\ref{subsec_bg}) we can describe our detection problem based on these two quantities $N_{\rm obs}/\theta$ and $N_{\rm obs}/\sigma_\mu$, together with the position of the peak in proper motion space. We call these the ``effective parameters''.  The latter has, however, less relevance compared to other properties, as already seen. 
 
  These quantities depend in turn on other astronomical parameters that characterize the system and its position with respect to the observer. But the successful detection of an \uf~depends only on a limited number of combinations of them, that is on the effective parameters. The importance of the effective parameters is that they reduce the dimensionality of the parameter space where we need to determine the boundaries of successful detection of our procedure.  In particular, while we describe our \ufs~by using 9 physical parameters, the detection of these systems depends only on 3 (and mainly 2) effective parameters.

Here we test that this concept is indeed correct. To evaluate the dependency of the detection limits and significance on these effective parameters, we have built a library of $2\,000$ \ufs~with varying $N_{\rm obs}$, but keeping constant the apparent sizes in the sky and proper motion planes, as well as projected center of mass velocity. Note that there is no straightforward way of generating \ufs~with the same real size in proper motion space $\sigma_\mu$ because of the effects of observational errors in proper motion (Section~\ref{subsec_planes}), and therefore we do it approximately by generating \ufs~with constant $\Delta\mu$, assuming that $\Delta\mu\sim\sigma_\mu$. Thus in this exercise we vary the physical parameters $r_{\rm h}$, $\sigma_V$ and $D$ in a way that their combination (Eqs.~\ref{e_lbsize} and~\ref{e_pmsize}) result in constant $\theta$ and $\Delta\mu$. We also change $V_{\rm gal}$ with distance in order to obtain the same proper motion peak for these \ufs.

\begin{figure}
\begin{center}
\centering

\includegraphics[width=0.43\textwidth]{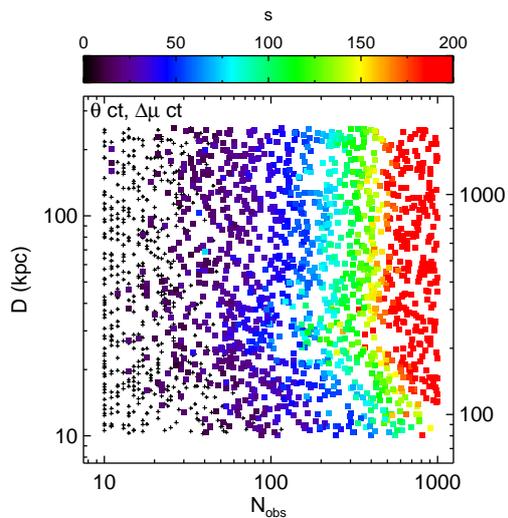}

\caption{Detectability tests run with a library of synthetic \ufs~built with constant apparent size in the sky $\theta$ and proper motion plane $\sigma_\mu$, as well as projected center of mass velocity. Regions of constant effective parameters $N_{\rm obs}/\theta$ and $N_{\rm obs}/\sigma_\mu$ are vertical lines in this plot. The panel show the significance $s$ as function of the physical parameters distance (\emph{left axis}) and half-light radii (\emph{right axis}) as a function of $N_{\rm obs}$. \ufs~with significance over 200 have been plotted with a colour saturated at this value. Black crosses indicate \ufs~that were not detected. }
\label{f_test60}
\end{center}
\end{figure}

Fig.~\ref{f_test60} shows the significance of the \ufs~of this experiment as function of $N_{\rm obs}$ and distance $D$ (left axis). The right axis shows the half-light radii $r_{\rm h}$ which is related to $D$ through Eq.~\ref{e_lbsize} to produce the same $\theta$. Although we do not include them in this plot, one could also draw other axes for $\sigma_V$ and $V_{\rm gal}$, which are also related to $D$ to produce the same $\Delta\mu$ and the same peak position in proper motion space. As $\theta$ and $\Delta\mu$ are constant for all these \ufs, the effective parameters $N_{\rm obs}/\theta$ and $N_{\rm obs}/\sigma_\mu\sim N_{\rm obs}/\Delta\mu$ are constant along 
 vertical lines in this plot. We see that equal significance contours are approximately vertical (but see discussion below),  which illustrates that indeed, for constant effective parameters the significance does not depend on the physical parameters ($D$, $\sigma_V$, $r_{\rm h}$) for these experiments (contrast this with the case of Fig.~\ref{f_tests}) but only on $N_{\rm obs}/\theta$ and $N_{\rm obs}/\sigma_\mu$. 

However, the colours do not follow exactly vertical contours. This is due to the effects of observational errors in proper motion which make the apparent size of the \ufs~in the proper motion plane $\sigma_\mu$ oscillate with $D$ as already explained in Section~\ref{subsec_planes} (Fig.~\ref{f_errorsize20}). This is the same effect as in Fig.~\ref{f_tests} (bottom panel). 
Here we generated these \ufs~with $\Delta\mu$ constant, but not $\sigma_\mu$.
In Fig.~\ref{f_test60} we can see how \ufs~at around $30\kpc$ are better detected because at this distance the apparent size $\sigma_\mu$ decreases. Although there is a similar effect around $70$ and $135\kpc$, these are not so clearly seen here.

%-------------------------------------

\subsection{Limits of detection and completeness }
\label{subsec_limits}

Having seen that \ufs~with the same effective parameters have the same significance, we can explore the detection limits and completeness as function of them alone. In the following we show the significance of a large synthetic library (15\,000 \ufs) that spans a large range in the effective parameters. All of the physical parameters of the \ufs~in this test are varied, except the LOS ($l$ and $b$).

\begin{figure}
\begin{center}
\includegraphics[width=0.42\textwidth]{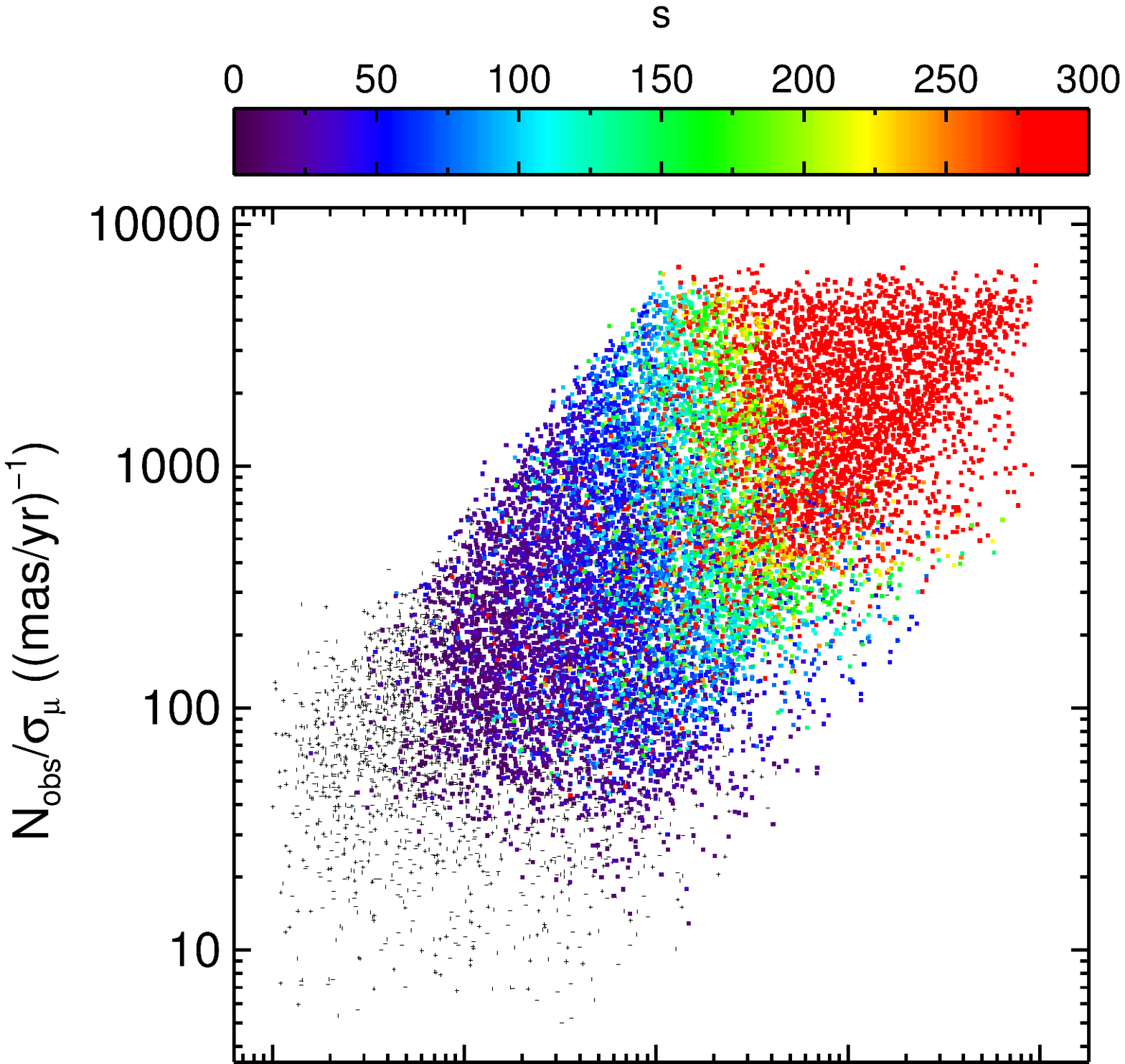}

\includegraphics[width=0.42\textwidth]{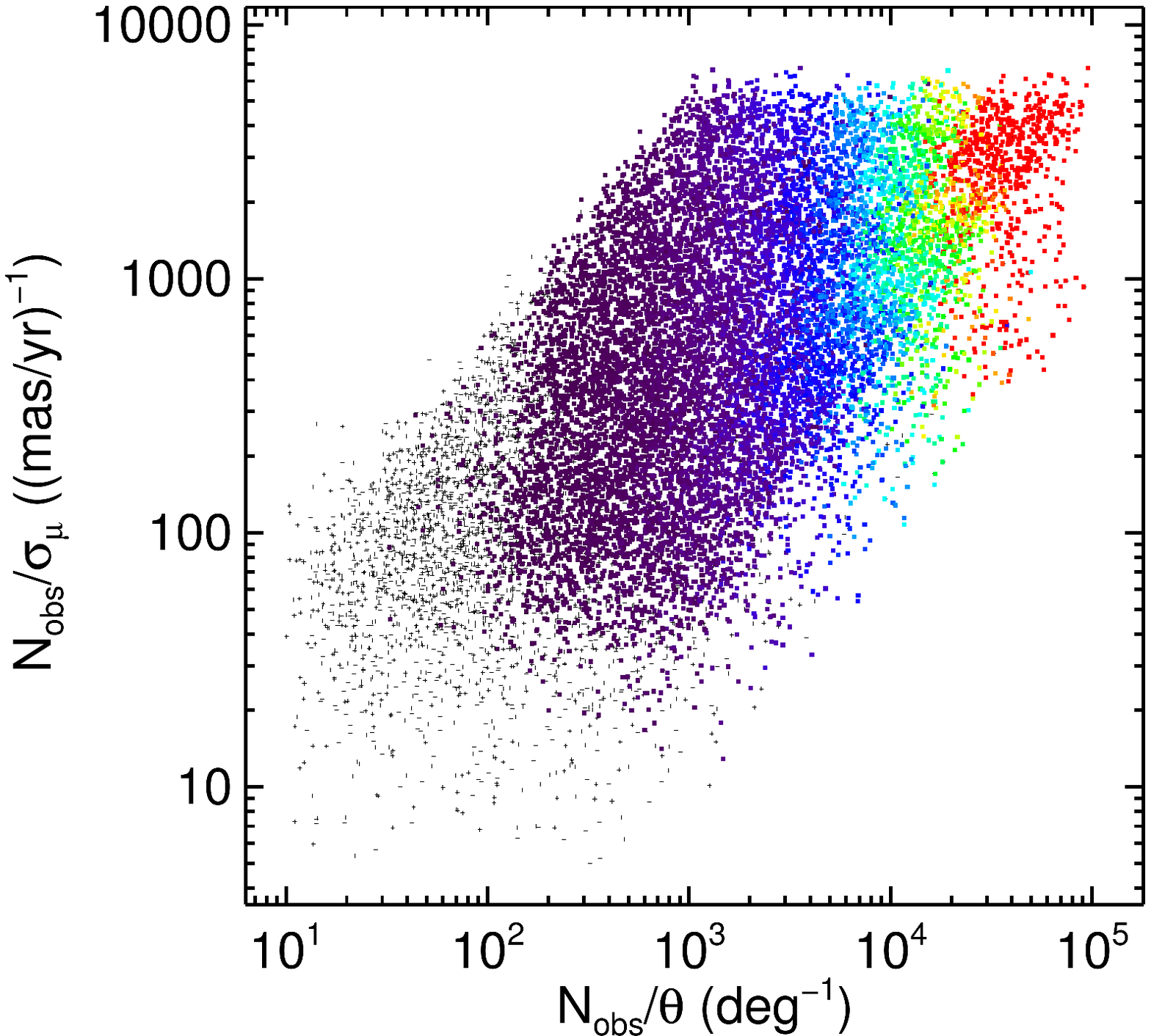}
\caption{\emph{Top:} Detectability tests run with a large library of synthetic \ufs~as function of two effective parameters: $N_{\rm obs}/\theta$ and $N_{\rm obs}/\sigma_\mu$. The color scale indicates the significance $s$. \ufs~with significance over 300 have been plotted with a colour saturated at this value. Black crosses indicate \ufs~that were not detected. \emph{Bottom:} Significance of the detections of the same library but being these only detections in the sky plane (see text for details).}
\label{d_test90}
\end{center}
\end{figure}

Fig.~\ref{d_test90} (top panel) shows the results of this test as function of the two effective parameters $N_{\rm obs}/\theta$ and $N_{\rm obs}/\sigma_\mu$. 
The colours are well separated, i.e. not strongly mixed, in this plot, showing that despite the physical properties of the \ufs~being very different\footnote{Remember that this is not a cross section as in Fig.~\ref{f_tests}.}, the significance of the detections depends mainly of these two effective parameters. Though the third effective parameter (position of the peak in proper motion space) changes in the \ufs~of this test, the uniformity of the colours in a certain region of this plot indicates that its influence is not as relevant as that of the other parameters, as already shown.

More in detail we can also see that for a higher fraction of the plot and specially the upper half, the colours follow a approximately vertical structure, i.e. the significance is mainly given by $N_{\rm obs}/\theta$. For the lower part, the contours are more curved. Finally, note also how the undetected objects lie in the regions of low $N_{\rm obs}/\theta$ and/or low $N_{\rm obs}/\sigma_\mu$, i.e. most diffuse objects.

We find that the minimum significance of our positive detections is $s\sim5$. This is because of the threshold imposed to $\ln P$ in order to filter false detections (see Section~\ref{subsec_threshold}).

\begin{figure}
\begin{center}
\includegraphics[width=0.42\textwidth]{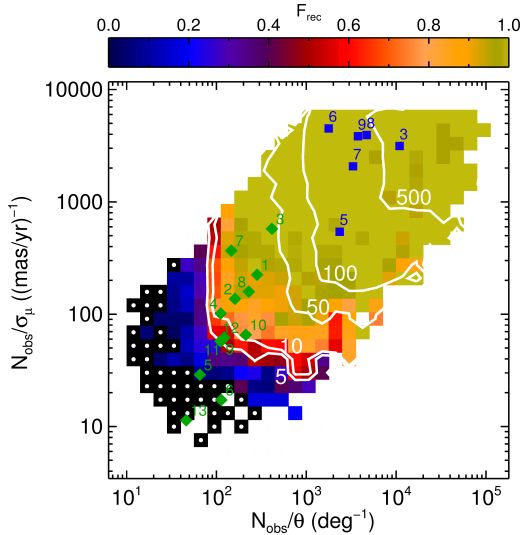}
\caption{Fraction $F_{\rm rec}$ of detected \ufs~as function of effective parameters. Black squares with a central white dot are regions where the fraction recovered is exactly equal to 0 (this is to differentiate from regions with small recovered fraction). The blue squares and the green diamonds show the estimated positions of classical dSphs and \ufs, respectively,
with labels as in Fig. \ref{f_rhsig}. White contours indicate the detection significance from Fig.~\ref{d_test90} (top).}
\label{f_test90}
\end{center}
\end{figure}

In Fig.~\ref{f_test90} we use the same test described above to estimate the fraction $F_{\rm rec}$ of detected \ufs~in each region of the effective parameter space. To do this, we have binned logarithmically this space and computed how many of the generated \ufs~in each bin are successfully detected. We only plot bins with at least 4 simulated \ufs. The median number of \ufs~in each cell is 14. 
%the maximum is 44.
Note how for most of the space explored this fraction is close to 1 (ocher colours). There is a transition zone where fractions go from $\sim0.5$ to $\sim0.2$. To differentiate regions with small recovered fraction from regions with this fraction equal to 0 (all of them with dark colours), we have marked the latter with a white central dot. The region where our method is not able to detect objects is the low $N_{\rm obs}/\theta$ and/or low  $N_{\rm obs}/\theta$, as expected. 

In this plot we superpose white contours indicating the significance $s$ of the detections from Fig.~\ref{d_test90} (top), computed as the median significance in the same grid used in this plot, including the cases that were not detected, that is with $s=0$, in the computation. Note that these contours are just approximate, given that there is some dispersion in the significance. For instance, around the contours of $s=5$ and $s=10$, the dispersion in $s$ is of 5-10. We see how the transition zone corresponds to the region of significance roughly around  $5$. 
For significance higher than 10, the recovered fractions is between  0.7 and 1.0. 

We also plot in Fig.~\ref{f_test90} the estimated values of effective parameters for the known MW satellites (classical dSph and \ufs, blue squares and green diamonds, respectively). The values of $\sigma_\mu$ for these know systems have been estimated by interpolating in a plot similar to Fig.~\ref{f_errorsize20} (bottom) but for velocity dispersions $\sigma_V$ between 5 and $10\kms$. As expected, all of the classical satellites of the MW lie in a region of effective parameters where our algorithm applied to the mock \gaia data successfully detects all simulated systems.  Note that some of the classical satellites lie outside the higher limits of the plot. Out of the 13 \ufs~that would have observable stars by \gaia (Leo T would not be observable),   1 of them lies in a region where $F_{\rm rec}$ is 1.0 (Boo (3)) and 4 of them lie in regions with recovery fraction of 0.9 (CVnI (1), Her (2), UMaII (7), CmB (8)). Besides, WilI (10), UMa (4), SegII (12) and BooII (9) are recovered with fractions of 0.8, 0.7, 0.6, 0.5, respectively. SegI (11) and LeoIV (5) are in regions with $F_{\rm rec}$ of 0.3 and 0.1, respectively. Finally, CVnII (6) and LeoV (5) are outside the limits of detection.
Note that in this plot we see how systems with effective parameters similar to the known \uf~would be detected by our algorithm. But some of the known \uf~have a small number of observables stars  (e.g. LeoIV (5) and SegII (12) have $N_{\rm obs}\sim5$ and $\sim7$, respectively) and our tests are done for a minimum of $N_{\rm obs}\sim10$. Nevertheless, for these low luminosity known cases, the estimated $N_{\rm obs}$ is very uncertain.

It is outstanding that our tests indicate that it is possible to detect with \gaia \ufs~similar to some of the ones detected by SDSS which is $\sim2$ mag deeper. This is because, whereas the \ufs~of SDSS were detected with photometry alone, our search is done also in the proper motion plane. The bottom panel of Fig.~\ref{d_test90} is the same as the top panel, but plotting the significance that would correspond to these detections if the search had been made only in the sky plane, that is not including proper motion data in the detection algorithm. The significance $s$ is now calculated through $s=({N_{\rm sky}-\langle N_{\rm sky}\rangle})/{\sqrt{\langle N_{\rm sky}\rangle}}$, where $N_{\rm sky}$ is the number of stars in the detected peak in the sky. We only plot the significance for the detections that had at least\footnote{Note that in the top panel the minimum $s$ found was 5.} $s=3$. The remaining simulated \ufs~are plotted with black crosses. The colours follow now  vertical contours as the vertical axis plays no role. Comparing this plot with the top panel we see how much the significance increases when the proper motions are included in the search. Red colours (maximum significance) are only achieved for higher $N_{\rm obs}/\theta$, that is for more densely populated objects in the sky (right part). Also the limits of detection are now located at larger $N_{\rm obs}/\theta$.

\begin{figure}
\begin{center}
\includegraphics[width=0.42\textwidth]{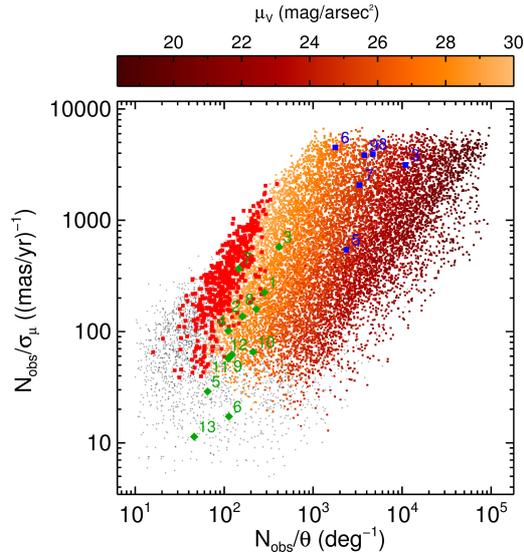}
\caption{Surface brightness of the detected \ufs~as function of the effective parameters. Black dots indicate \ufs~that were not detected. Detected objects with surface brightness larger than 30 mag/arsec$^2$ are highlighted in red colours. The blue squares and the green diamonds show the estimated positions of classical dSphs and \ufs, respectively, with labels as in Fig. \ref{f_rhsig}.}
\label{mu_test100}
\end{center}
\end{figure}

Fig.~\ref{mu_test100} shows the surface brightness of the detected \ufs~as function of the effective parameters. Contours of similar surface brightness are approximately diagonal in this parameter space. We have marked with red the {\em detected} synthetic \ufs~with surface brightness dimmer than 30 mag/arsec$^2$ which is the global SDSS surface brightness limit as found in \citet{Koposov08}. Very interestingly, the red squares mark out an area in the parameter space of \ufs~less bright than the SDSS limit and that would be possible to explore with {\em Gaia}. Note, nonetheless, that this region has a recovery fraction $F_{\rm rec}$ smaller than 0.8. 

In Fig.~\ref{mu_test100} (and in Fig.~\ref{f_test90}) all known satellites of the MW lie in an approximate diagonal line in the effective parameter space. We believe that this is a projection of the fundamental curve mass-radius-luminosity studied in e.g. \citet{Tollerud2011}, or more in detail, a consequence of the Faber-Jackson and the $r_{\rm h}$-$L$ scaling relations. The plot shows that our algorithm would be able to detect objects that are outside this diagonal. However, \ufs~that lie below the diagonal have surface brightness brighter than the SDSS limit and they would have already been detected, unless they are all located outside the SDSS footprint. On the other hand, part of the red region with surface brightness dimmer than the SDSS limit but that \gaia could probe lies outside the diagonal and, therefore, the detection of objects in it relies on the existence of them. Note, however, that the scatter across the diagonal is large.

\begin{figure}
\begin{center}
\includegraphics[width=0.4\textwidth]{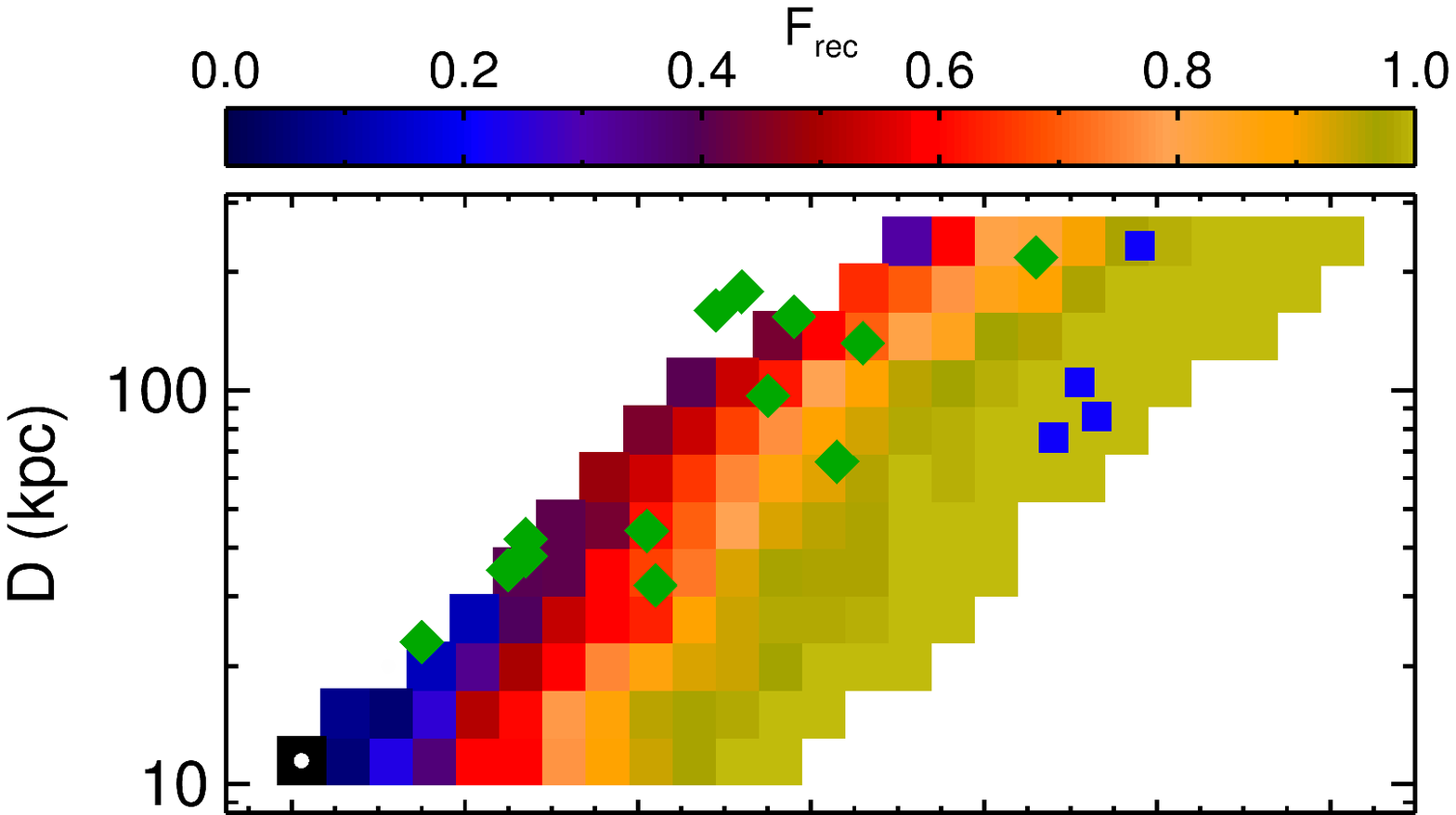}

\includegraphics[width=0.4\textwidth]{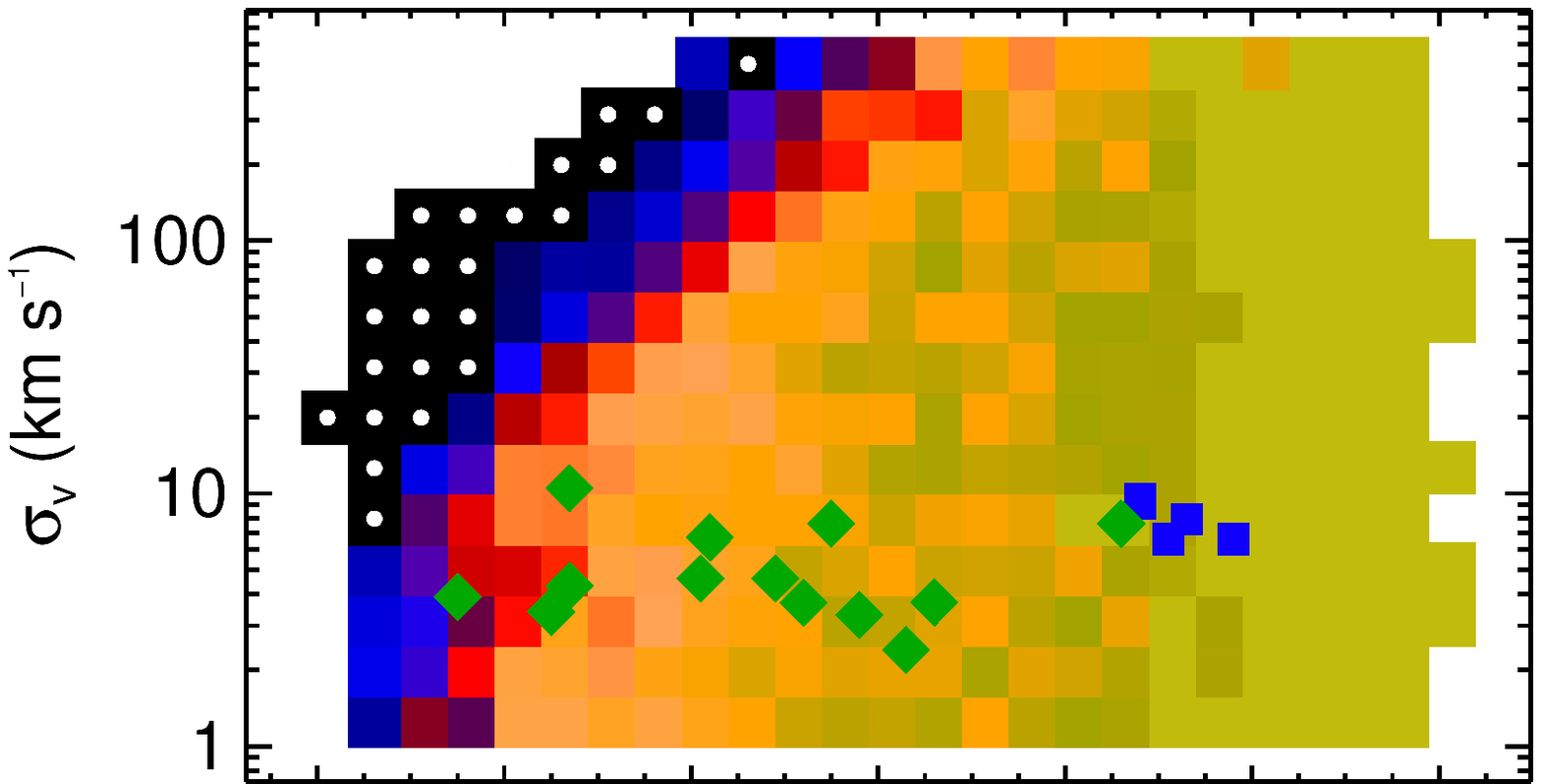}

\includegraphics[width=0.4\textwidth]{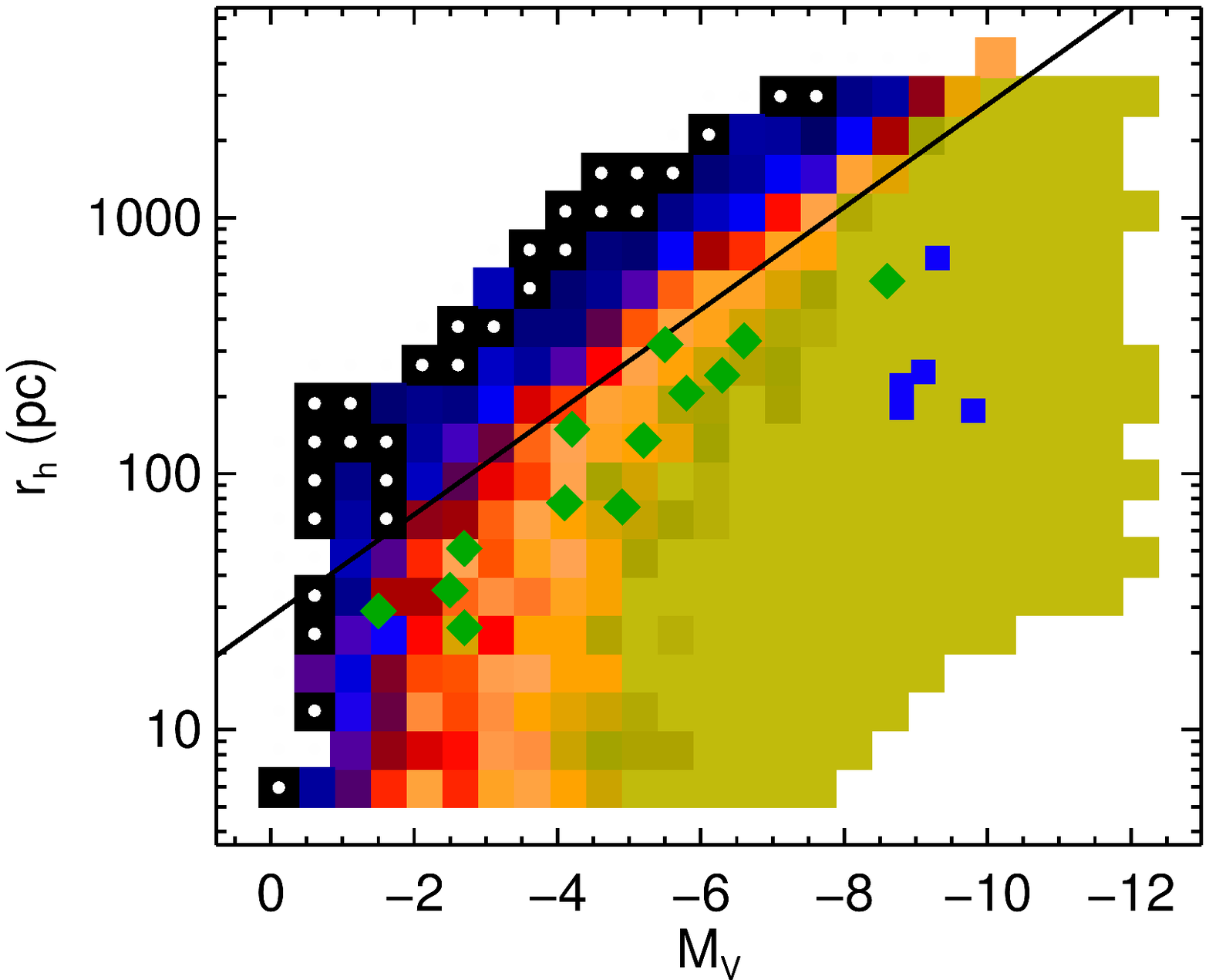}
\caption{Fraction $F_{\rm rec}$ of detected \ufs~(colour-scale) as a function of $D$ (\emph{top}), $\sigma_V$ (\emph{middle}) and $r_{\rm h}$ (\emph{bottom}) versus $M_V$. Black squares with a central white dot are regions where the fraction recovered is exactly equal to 0. The blue squares and the green diamonds show the estimated positions of classical dSphs and \ufs, respectively. The black diagonal line in the bottom panel shows the SDSS surface brightness limit $\mu_V=$ 30 mag/arsec$^2$.}
\label{f_test901}
\end{center}
\end{figure}

\begin{figure}
\begin{center}

\includegraphics[width=0.4\textwidth]{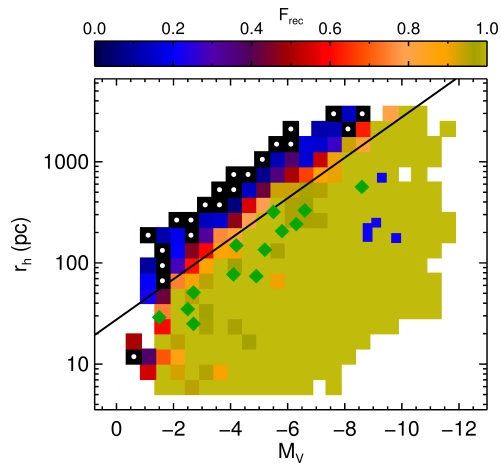}
\caption{Same as low panel of Fig.~\ref{f_test901} but only considering \ufs~with velocity dispersion $\sigma_V<10\kms$.}
\label{f_test901_2}
\end{center}
\end{figure}

Fig.~\ref{f_test901} illustrates the recovered fraction (colour-scale) of \ufs~but now in terms 
of the physical parameters $D$, $\sigma_V$ and $r_{\rm h}$ as a function of $M_V$. In these panels, the region with fractions between $0.9$ and $1$ (ocher colours) occupies a much smaller portion of the explored ranges. This is because, in these plots, in any given bin only two physical parameters are fixed while the remaining are varying in the entire explored range, which can result in a very different detection significance.
This has the effect of lowering $F_{\rm rec}$ on average, while also resulting in more diffuse boundaries between areas with different $F_{\rm rec}$, as opposed to the sharp boundaries seen in the effective parameter plane of Fig. \ref{f_test90}, which corroborates the fact that our detection scheme does depend mainly on the effective parameters. In this figure we also show the positions of known \ufs~and classical dSph galaxies, though without the number labels.
These are shown to illustrate the typical recovery fraction one would expect for a galaxy with, e.g. a given $r_h-M_V$, if its velocity dispersion and other parameters are unknown but restricted to the range spanned by our library.

This implicit dependence on the other physical parameters means that the behaviour of $F_{\rm rec}$ in these plots will change depending on the assumed distributions for the different parameters, and so, strictly speaking the reported $F_{\rm rec}$ is only valid under the assumed log-uniform distributions. For instance, if we consider only \ufs~with small velocity dispersion ($\sigma_V<10\kms$; Fig.~\ref{f_test901_2}), the boundaries of detection improve significantly, i.e. the algorithm could detect larger \ufs~at the same given $M_V$. The simple distributions assumed for the physical parameters do allow us, however, to illustrate the limits of our method. Finally, it is worth noticing that the recovered fractions shown in Figs.~\ref{f_test90} and \ref{f_test901} can be interpreted in a probabilistic sense as the probability that any individual galaxy is detected by our method, given two of its physical or effective parameters.

 The diagonal line in the bottom panel of Fig.~\ref{f_test901} shows the SDSS surface brightness limit of $\mu_V=$30 mag/arcsec$^2$. Lines of equal surface brightness are diagonal lines with slope of 5 in this plot. The detection limits of our algorithm (for example considering the line delineated by the red or blue coloured bins) have a similar slope at a slightly lower surface brightness with an additional vertical limit at $M_V\sim-1.5$ 
(but note that all these depends on the underlying distribution of physical parameters).

The shape of the detection frontiers in the lower panel of Fig.~\ref{f_test901} is similar to the ones of \citet{Koposov08} in their Figs.~10 and 11. For brighter systems the detection limits follows a diagonal line with the slope of a constant surface brightness line, followed by a vertical cut at certain absolute magnitude. In the case of \citet{Koposov08}, the surface brightness and absolute magnitude limits vary as function of distance. However, the detection limits of the two studies are not directly comparable because our method is based on different information, as it includes kinematics. Our effective parameter space, where the detection limits are defined, is essentially different (with more dimensions).

%--------------------
\subsection{Changing the background}
\label{subsec_bg}

The results of the previous section correspond to our fiducial field $(l,b)=(90^\circ,30^\circ)$. We now explore how these results change for different LOS's that have a different number of stars $N_{\rm BG}$ and a different distribution of proper motions. 

Let $s$ be the significance of a detection in the fiducial field, in which we have an \uf~with $N_{\rm obs}$ and a background with $N_{\rm BG}$ observed stars. Because $N_{\rm BG,in}\propto N_{\rm BG}$, from Eq.~\ref{eq_sr}, it follows that $s\propto {N_{\rm obs}}/{\sqrt{N_{\rm BG}}}$. Then, given an \uf~with the same effective parameters but in an arbitrary LOS, for which the number of background stars $N'_{\rm BG}$ has changed in a ratio $r=N'_{\rm BG}/N_{\rm BG}$, its significance is 
\begin{equation}\label{e:s_factor2}
s'\sim\frac{s}{\sqrt{r}}
\end{equation}
This is a useful relation that allows us to establish the significance and detection limits in the effective parameter space for different LOS's without running additional experiments. 

For instance, the number of background stars in the fiducial field is $N_{\rm BG}=1413$ and for two different LOS's at $(l,b)=(90^\circ,55^\circ)$ and $(l,b)=(90^\circ,80^\circ)$ this is  $N_{\rm BG}=525$ and $N_{\rm BG}=377$, respectively. Therefore the background has decreased by factors $r=0.37$ and $r=0.27$, respectively, with respect to the fiducial case. Thus, we expect the significance of \ufs~with the same effective parameters to increase by $s'=1.6s$ and $s'=1.9s$, respectively.

In the following we check that this relation is correct. We use a library of 1000 \ufs~(randomly extracted from the library of Section~\ref{subsec_limits}), and locate copies of it in different LOS's. We then compare the one-by-one significance for different pairs of LOS's. Note, however, that because we keep the proper motion of each \uf~constant, its relative position with respect to the centroid will change depending on the LOS (because the underlying distribution changes), thus changing one of the effective parameters. For this reason, and also because of the approximations used to derive Eq.~\ref{e:s_factor2} and that \gaia errors change with LOS, we expect a certain dispersion around the values predicted by Eq.~\ref{e:s_factor2}.

\begin{table}
 \setlength{\tabcolsep}{1.5pt}
\caption{Comparison of the significance of an ensemble of 1000 \ufs~located at different fields.}\label{t_bg}
\centering
\begin{tabular}{ccccccccc}\hline\hline
       \multicolumn{2}{c}{field A} & \multicolumn{2}{c}{field B}  & expected & median& AMD & $P_{10}$   &  $P_{90}$ \\
        $l$  & $b$&$l$& $b$   &  $s_B/s_A$&  $s_B/s_A$&  &   &  \\
           \hline\hline
 90  &   30  &   90  &   42  &    1.4  &    1.2  &    0.2  &    0.8  &    1.9  \\
 90  &   30  &   90  &   55  &    1.6  &    1.4  &    0.3  &    0.9  &    2.4  \\
 90  &   30  &   90  &   68  &    1.9  &    1.7  &    0.4  &    1.0  &    3.0  \\
 90  &   30  &   90  &   80  &    1.9  &    1.9  &    0.4  &    1.0  &    3.2  \\
 
 90  &   42  &   90  &   55  &    1.2  &    1.2  &    0.2  &    0.7  &    1.8  \\
% 90  &   42  &   90  &   68  &    1.4  &    1.4  &    0.3  &    0.9  &    2.3  \\
% 90  &   42  &   90  &   80  &    1.4  &    1.5  &    0.3  &    0.9  &    2.3  \\
 
170  &   30  &  170  &   42  &    1.3  &    1.2  &    0.2  &    0.8  &    1.8  \\
170  &   30  &  170  &   55  &    1.3  &    1.4  &    0.2  &    0.9  &    2.2  \\
170  &   30  &  170  &   68  &    1.4  &    1.4  &    0.2  &    0.9  &    2.1  \\
170  &   30  &  170  &   80  &    1.4  &    1.5  &    0.3  &    1.0  &    2.4  \\

 90  &   30  &  170  &   30  &    1.4  &    1.4  &    0.5  &    0.6  &    3.3  \\
 90  &   30  &  170  &   42  &    1.8  &    1.7  &    0.6  &    0.7  &    4.1  \\
 90  &   30  &  170  &   55  &    1.9  &    1.9  &    0.7  &    0.8  &    4.4  \\
 90  &   30  &  170  &   68  &    2.0  &    1.8  &    0.7  &    0.8  &    4.5  \\
 90  &   30  &  170  &   80  &    2.0  &    2.0  &    0.8  &    0.9  &    4.5  \\

\hline
\end{tabular}
\end{table}

Table~\ref{t_bg} compares the expected value of $s_B/s_A$ with the median observed values computed with all 1000 \ufs~at different pairs of LOS A and B. We also give the Absolute Median Deviation (AMD), and the 10 per cent ($P_{10}$) and 90 per cent ($P_{90}$) percentiles. The median ratios $s_B/s_A$ differ at most 0.2 from the expected values. We see also that there is some expected dispersion with respect to this value. The cases where more dispersion is observed are when we compare fields at different longitudes (last rows). This is because in these cases the distribution of background proper motion changes the most. However, by looking at the percentiles we see that most of the dispersion comes from values that are higher than the expected value (thus improving the significance). The $P_{10}$ is always around 1. This means that the significance of all the fields B is smaller than predicted, but larger than the significance of the fields A in $\sim40$ per cent of the cases. For $\sim50$ per cent of the cases the significance of all the fields B is larger than expected.

In conclusion, the significance of the detections of our fiducial field at $l=90\deg$ and $b=30\deg$ are maintained or improved in $\sim90$ per cent of the simulated cases in the other fields where $N_{\rm BG}$ was smaller. The boundaries of the detection will also improve for these fields. But the scaling between significance and fraction of recovery is not straightforward and one would need to evaluate this in each particular field. Eq.~\ref{e:s_factor2} offers, however, a fast approximate way of comparing the success of the detections in different LOS's.

%------------------------------------------------------------------------------------------------------------------

\section{Caveats}
\label{sec_caveats}

The method that we have introduced here has its limitations and assumptions, which we will review here.

First of all, we emphasize that our method does not aim to characterize and study \ufs~but it is a probabilistic method to identify possible candidates. When applied to real data, it will provide us with a list of candidates that will need to be studied in detail. The colour-magnitude diagrams 
of \gaia photometry can be used for this, as well as to derive morphological properties, kinematics, distances, etc, once a proper filter to select the \uf~population is designed, as has 
been done with SDSS \citep{willman02}. Also, a follow up using ground-based facilities will be required to obtain radial velocities and detailed chemical abundances.

Our procedure has been tested for Galactic latitudes above $30\deg$ in fields of $2\deg\times 2\deg$ and has been designed for \ufs~at distances larger than 10 kpc. If we want to apply it to search for nearer systems, different parallax and $\log g$ cuts would be needed. However, larger scales in the sky would have to be probed, increasing the background level. Another limitation of the method is that it is optimized for \ufs~with angular sizes smaller than the $2\deg\times 2\deg$ fields, and the detection of larger systems would require, again, to probe larger scales in the sky and perhaps a different strategy.

Likewise, it has been tested for \ufs~modeled as Plummer spheres with isotropic velocity distributions where light follows mass. A change in these assumptions that results on a variation in the footprints in the sky or proper motion planes, will change the effectiveness of our method, 
although the limits we have encountered should remain the same, when expressed in terms of the effective parameters. It is the mapping from structural to effective parameters that would 
need to be established for the new \uf~models.  Similarly, the boundaries that define the limits of our detection method in $N_{obs}$ will simply map into different boundaries in stellar luminosity, if the stellar population content is changed.

Our background clearly comes from smooth distributions without streams or clouds. A clumpy halo may affect the number of background stars compared to our estimations with GUMS. However, our algorithm will also detect other systems that are not necessarily \ufs~as long as they present some coherence in the 4-dimensional space that we use. These detections, rather than being considered as additional false positives,  will be interesting systems to be followed-up. 

Previous studies to detect \ufs~with photometric surveys apply an isocrhone masking or a probabilistic modelling in the colour-magnitude or colour-colour diagram in order to filter out field stars \citep[e.g.][]{Koposov08}. Instead, here we do cuts using parallax and surface gravity. Some preliminary tests show that the addition of an isocrhone masking in the Gaia $G$ vs. $G_{\rm{BP}}-G_{\rm{RP}}$ plane in our algorithm may be beneficial in particular cases. This merits a separate investigation that we aim to undertake in the future.

Although we have not included unresolved galaxies and quasars in our simulated background, we have checked that these will have a minor effect in our results. According to \citet{BailerJones13} (their table 3) the fraction of misclassified galaxies and quasars is 2.5\% (2\% misclassified as stars and 0.5\% as binary systems) and 8.9\% (5.9\% as stars and 0.1\% as binary systems), respectively. From GUMS simulations we have estimated that the number of galaxies and quasars in our fiducial field would be 4051 and 120, respectively. Therefore, there will be $\sim110$ objects (mainly galaxies) misclassified as stars, corresponding to an increase of 7\% of the back/foreground population in our fiducial field, and up to $\sim30$\% for other LOS ($(l,b)=(180^\circ,80^\circ)$). However, this increase in the number of field stars will imply approximately a change in the significance of the detections only of a factor 0.96 and 0.88 (Eq.~\ref{e:s_factor2}), respectively, in the two LOS described.

One important aspect of the \gaia astrometric data that we have not taken into account in this work is the issue of covariances in the estimated astrometric parameters. As explained in \cite{Lindegren2012}, the statistical correlation between the different astrometric parameters will occur between the parameters of the same source and also between the parameters of different sources. The within-source error covariances can be similar for collections of sources in small areas of the sky, as can be seen for example in the statistical plots in Volume 1 to the Hipparcos Catalogue documentation \citep{ESA1997}. In the proper motion plane for small areas on the sky (such as used in this study) this can lead to apparent structure in the proper motion distribution (caused by elongated and preferentially oriented error-ellipses).The between-source covariances will have a similar effect and are estimated in the case of \gaia to be most pronounced over areas of order $0.3^\circ$ radius on the sky \citep[the value of the correlation half-length estimated in][]{Holl2010}. This means that for a large fraction of \ufs~the between-source correlations will be important in addition to the within-source correlations. To first order the main effect will be that the interpretation of the WT 
maps will be more involved, where a distinction will have to be made between real and spurious structure in the proper motion plane.

The within-source covariance matrix of the astrometric parameters will be provided as part of the \gaia data releases. The covariance matrix of the astrometric parameters of different sources cannot be calculated for the full \gaia catalogue but it is feasible to do so for limited groups of sources as demonstrated in \cite{Holl2012a} and \cite{Holl2012b}. Hence we will be 
able to account for the error covariances but we defer to a future study the details of how to implement this in practice.

%------------------------------------------------------------------------------------------------------------------

\section{Discussion and conclusions}
\label{sec_conc}

We have introduced an automatic procedure to identify \uf~candidates in the future \gaia database and charted its detection limits. The main advantages of using \gaia data on the search of \ufs~are, first, the inclusion of kinematics (proper motions) in the detection algorithm for the first time; and second, the \gaia full sky coverage, being the first unbiased homogeneous survey to be used for this purpose.

Our procedure identifies significant overdense peaks in the planes of the sky and of proper motions that share common stars. Then the probability of this occurring by chance is assessed and used to discard spurious detections. We have used a library of $\sim 30\,000$ synthetic \ufs~to probe the 9-D space of intrinsic ($L_V$, $r_h$, $\sigma_V$) and extrinsic ($l$, $b$, $D$, $V_{\rm gal}$, $\phi_V$, $\theta_V$) \uf~parameters, spanning ranges that extend well beyond those occupied by currently known systems. 

We have identified the ``effective parameters'' that our algorithm depends mainly on. The main two are the ratios of the number of observable stars by \gaia in the \ufs~to their apparent sizes in the sky ($N_{obs}/ \theta$) and proper motion planes ($N_{obs}/ \sigma_\mu$). The position of the peak in proper motion with respect to the background also influences the 
detection, but is not as relevant. These parameters reduce the dimensionality of our problem to 3, mainly 2, parameters.

We have charted the limits of detectability and completeness (recovery fraction) of our search in the effective parameter space (Fig.~\ref{f_test90}) for a LOS at $l=90\deg$ and $b=30\deg$. Detections can be made with high significance over most of the explored region, which includes the majority of the currently known \ufs, with a recovered fraction that remains above 70 per cent over most of it. It is only in the corner of small effective parameters that the efficacy of our method decreases abruptly. On the other hand, the limits of our detection procedure can not be described in terms of a limiting surface brightness alone (Fig.~\ref{mu_test100}), because of the inclusion of kinematics in the search.

%  The variation as a function of distance of the dominant type of star observable by Gaia (e.g. HB, MSTO, RGB) in an \uf, introduces a slight modulation in the detectability (Fig. 10), due to a variation of the proper motion errors. 

%????????????
%In the full range of $\sigma_V$ explored ($\sigma_V<400$ \kms), the recovery fraction is $\sim80$\% at the surface brightness limit of SDSS for \uf brighter than $M_V=-3$. For \ufs~brighter than $M_V=-5$ it is close to $100$\%, even for half-light radii larger than those of known systems. Note that the range explored includes systems with $\sigma_V$ up to 40 times larger than that of known satellites. 

We have derived a relation that allows us to know the approximate detection significance of the synthetic \ufs~at  LOSs with a different number of background stars. The translation from significance to recovery fraction is not straightforward and one would need a more thorough characterization per LOS. However, most of the results presented here are for a pessimistic case compared to higher latitudes, or to the outer galaxy ($l=180\deg$), where we expect less field contamination.

Furthermore, we have explored the extent to which current detectability limits can be pushed forward, opening the possibility of detecting real systems hitherto not found. We have found that there is a region in the effective parameter space where there are currently no observed systems. Part of this region corresponds to \ufs~with surface brightness brighter than the SDSS limit and, therefore, they would have already been detected, unless they are all located outside of the SDSS footprint. But more interestingly, we have seen that \gaia will be able to probe a region of the effective parameter space of surface brightness dimmer than the SDSS limit, if such objects exist, albeit with a recovery fraction smaller than 0.8. Note that the recent \uf~discoveries made with DECam have similar surface brightness to the ones detected by SDSS \citep[see Fig. 17 in][]{Koposov2015}. Also because of the different detection methodologies followed by SDSS and DECam compared to Gaia, the nature of the detection limits is completely different, thus offering the possibility to explore uncovered regions of the parameter space (both with respect other surveys in the north and in the south) and for all sky.

%%%%%%%%%%%%%%%%%%%%%%%%%%%%%%%%%%%

We can make a very rough estimation of the number of \ufs~that \gaia will detect from the recovery fractions that we have found for our synthetic search (Fig.~\ref{f_test90}), assuming isotropy on the distribution of satellites in the MW halo, and considering only SDSS \ufs.
There is 1 known object (Boo) that would be detected with a recovery fraction of 1.0, 4 objects (CVnI, Her, UMaII, CmB) with a  fraction of 0.9, and 4 objects (WilI, UMa, SegII, BooII) with fractions of 0.8, 0.7, 0.6 and 0.5, respectively. We do not count objects with a recovery fraction below 0.5. This makes a total of 7.2 \ufs~in a sky area equivalent to SDSS ($\sim1/5$ of the sky, \citealt{Koposov08}). If we assume that \gaia will detect \ufs~only above $b=30\deg$,  which corresponds to 1/2 of the sky, there should be of the order of $\sim10$ new \ufs~(i.e. currently not known) over the 1/2-1/5=3/10 of the sky that remains unexplored, that is subtracting the area already covered by SDSS. These calculations are based on the field at $l=90\deg$ and $b=30\deg$ but could be slightly better for higher latitudes.

However, by the arrival time of the Gaia catalogue (see below) other surveys such as ATLAS \citep{Shanks2013}, Pan-STARRS \citep{Kaiser2010} and DES \citep{Diehl2014} will have covered great fraction of this area. But a part of the South Galactic cap will still remain completely unexplored (at declinations below stripe SPT of DES). We estimate this to be a fraction of $\sim0.0195$ of the whole sphere (by taking the part of the spherical cap in equatorial coordinates below $\delta<-65\deg$ that lies in the range $\alpha\sim[-60, 90]$). Therefore, there should be of the order of $\sim1$ new \uf~in this unexplored area. However, we emphasize that our method uses information not used in other searches, namely proper motions, and thus, it could lead to new discoveries, made possible, not by the covered region in the sky or depth of probing, but by their motion in the sky.  As such, our method complements present searches.

Moreover, the number of discovered new candidates could be higher because as discussed above, \gaia could also detect more \ufs~with lower surface brightness than the SDSS limit. Besides, under the assumption of anisotropy in the spatial distribution of satellites, this number could be larger if the \gaia footprint happens to cover preferential directions. In fact, the importance of having a full-sky catalogue in this type of search for the first time is that it will allow to put constraints on the isotropic distribution of the satellites and, therefore, their origin.

 The known \ufs~with high recovery fraction mentioned above could be seen as standard systems for future \gaia discoveries but only in terms of effective parameters. Thus, one can not interpret this as if, for instance, all objects with the same half-light radius and the same distance as Boo will be detected, but rather as systems for which the combination of \emph{ all} physical parameters produce similar effective parameters will be detected with high probability. Note also that we have not considered in this calculation the influence of the third effective parameter, which we have shown to be less important.
%-----

Our proposed method can be applied fully to the third \gaia data release scheduled for\footnote{\url{http://www.cosmos.esa.int/web/gaia/release}} 2017/2018. This release will include the five-parameter astrometric solutions as well as the object classification (necessary to eliminate contaminant extra-Galactic objects) and astrophysical parameters such as $\log g$, necessary for filtering out foreground dwarfs. Preliminary searches could be conducted using earlier releases; e.g. with the first data release in summer 2016, using only on sky coordinates; or with the second data release in early 2017, using full sky and proper motion information, yet without the possibility of using the foreground filters as explained here, since astrophysical parameters will not yet be available.

Finally, there is the future possibility that the \gaia magnitude limit will be pushed down to $G=20.7$. This will obviously be positive in terms of the number of observable stars in each \uf, but will also increase the foreground/background contamination, so the effect in the detection probabilities will have to be assessed.

%\Teresa{An additional advantage of the availability of proper motions, apart from their inclusion on the search algorithm, is the possibility of studying the global proper motion of the \ufs~at later stage, when the candidate is confirmed \citep{Jin2015}} and \Octavio{perhaps, allows for the study of disruption signatures in the kinematics of these objects.}\Teresa{not sure about the later one}
%\Octavio{In the optimistic case that there are many detections, it will be possible to verify the correlations (fundamental curve) of galaxy parameters and the scatter}\Teresa{Not sure we have to enter in this discussion here}

%----------------------------------------------------------------------------------------------------------------

\section*{Acknowledgments}
We thank V. Belokurov, the referee, for a thorough review and insightful comments on the original manuscript. TA is supported by an ESA Research Fellowship in Space Science. CM acknowledges support from a post-doctoral fellowship of DGAPA-UNAM, Mexico.
CM and LA acknowledge support from DGAPA/UNAM grant IG100115 and the hospitality and support of the University of Barcelona during part of this investigation. FF acknowledge support from  MINECO (Spanish Ministry of Economy) - FEDER through grant AYA2012-39551-C02-01 and ESP2013-48318-C2-1-R, and from the European Community's Seventh Framework Programme (FP7/2007-2013) under grant agreement GREAT-ITN FP7 264895.  We thank the CEA and Nice Observatory for the MR software. We thank Merce Romero-Gomez and GaiaUB team for the code to simulate Gaia errors.

\bibliographystyle{mn2e}

% definiciones solo para el MNRAS para ApJ no ponerlas
\def\apj{ApJ}
\def\apjl{ApJ}
\def\aj{AJ}
\def\mnras{MNRAS}
\def\aa{A\&A}
\def\nat{nat}
\def\araa{ARA\&A}
\def\aap{A\&A}
\def\pasp{PASP}% Publications of the Astronomical Society of the Pacific
\def\apss{Ap\&SS}             % Astrophysics and Space Science

\bibliography{mybib}

\begin{thebibliography}{60}
\expandafter\ifx\csname natexlab\endcsname\relax\def\natexlab#1{#1}\fi

\bibitem[{{Anscombe}(1948)}]{Anscombe48}
{Anscombe} F.~J., 1948, {The Transformation of Poisson, Binomial and
  Negative-Binomial Data}, Vol.~34. pp. 246--254

\bibitem[{{Antoja} {et~al}\mbox{.}(2008){Antoja}, {Figueras}, {Fern{\'a}ndez},
  \& {Torra}}]{Antoja08}
{Antoja} T., {Figueras} F., {Fern{\'a}ndez} D., {Torra} J., 2008, \aap, 490,
  135

\bibitem[{{Antoja} {et~al}\mbox{.}(2012){Antoja}, {Helmi}, {Bienayme},
  {Bland-Hawthorn}, {Famaey}, {Freeman}, {Gibson}, {Gilmore}, {Grebel},
  {Minchev}, {Munari}, {Navarro}, {Parker}, {Reid}, {Seabroke}, {Siebert},
  {Siviero}, {Steinmetz}, {Williams}, {Wyse}, \& {Zwitter}}]{Antoja12}
{Antoja} T. {et~al.}, 2012, \mnras, 426, L1

\bibitem[{{Arenou}(2011)}]{Arenou2011}
{Arenou} F., 2011, in American Institute of Physics Conference Series, Vol.
  1346, American Institute of Physics Conference Series, {Docobo} J.~A.,
  {Tamazian} V.~S., {Balega} Y.~Y., eds., pp. 107--121

\bibitem[{{Bailer-Jones} {et~al}\mbox{.}(2013){Bailer-Jones}, {Andrae},
  {Arcay}, {Astraatmadja}, {Bellas-Velidis}, {Berihuete}, {Bijaoui},
  {Carri{\'o}n}, {Dafonte}, {Damerdji}, {Dapergolas}, {de Laverny},
  {Delchambre}, {Drazinos}, {Drimmel}, {Fr{\'e}mat}, {Fustes},
  {Garc{\'{\i}}a-Torres}, {Gu{\'e}d{\'e}}, {Heiter}, {Janotto}, {Karampelas},
  {Kim}, {Knude}, {Kolka}, {Kontizas}, {Kontizas}, {Korn}, {Lanzafame},
  {Lebreton}, {Lindstr{\o}m}, {Liu}, {Livanou}, {Lobel}, {Manteiga},
  {Martayan}, {Ordenovic}, {Pichon}, {Recio-Blanco}, {Rocca-Volmerange},
  {Sarro}, {Smith}, {Sordo}, {Soubiran}, {Surdej}, {Th{\'e}venin},
  {Tsalmantza}, {Vallenari}, \& {Zorec}}]{BailerJones13}
{Bailer-Jones} C.~A.~L. {et~al.}, 2013, \aap, 559, A74

\bibitem[{{Bechtol} {et~al}\mbox{.}(2015){Bechtol}, {The DES Collaboration},
  {Drlica-Wagner}, {Balbinot}, {Pieres}, {Simon}, {Yanny}, {Santiago},
  {Wechsler}, {Frieman}, {Walker}, {Williams}, {Rozo}, {Rykoff}, {Queiroz},
  {Luque}, {Benoit-Levy}, {Bernstein}, {Tucker}, {Sevilla}, {Gruendl}, {da
  Costa}, {Fausti Neto}, {Maia}, {Abbott}, {Allam}, {Armstrong}, {Bauer},
  {Bernstein}, {Bertin}, {Brooks}, {Buckley-Geer}, {Burke}, {Carnero Rosell},
  {Castander}, {D'Andrea}, {DePoy}, {Desai}, {Diehl}, {Eifler}, {Estrada},
  {Evrard}, {Fernandez}, {Finley}, {Flaugher}, {Gaztanaga}, {Gerdes},
  {Girardi}, {Gladders}, {Gruen}, {Gutierrez}, {Hao}, {Honscheid}, {Jain},
  {James}, {Kent}, {Kron}, {Kuehn}, {Kuropatkin}, {Lahav}, {Li}, {Lin},
  {Makler}, {March}, {Marshall}, {Martini}, {Merritt}, {Miller}, {Miquel},
  {Mohr}, {Neilsen}, {Nichol}, {Nord}, {Ogando}, {Peoples}, {Petravick},
  {Plazas}, {Romer}, {Roodman}, {Sako}, {Sanchez}, {Scarpine}, {Schubnell},
  {Smith}, {Soares-Santos}, {Sobreira}, {Suchyta}, {Swanson}, {Tarle},
  {Thaler}, {Thomas}, {Wester}, \& {Zuntz}}]{Bechtol2015}
{Bechtol} K. {et~al.}, 2015, ArXiv, astroph/1503.02584

\bibitem[{{Behroozi}, {Wechsler} \& {Wu}(2013){Behroozi}, {Wechsler}, \&
  {Wu}}]{Behroozi2013}
{Behroozi} P.~S., {Wechsler} R.~H., {Wu} H.-Y., 2013, \apj, 762, 109

\bibitem[{{Belokurov}(2013)}]{Belokurov2013}
{Belokurov} V., 2013, NewAR, 57, 100

\bibitem[{{Belokurov} {et~al}\mbox{.}(2008){Belokurov}, {Walker}, {Evans},
  {Faria}, {Gilmore}, {Irwin}, {Koposov}, {Mateo}, {Olszewski}, \&
  {Zucker}}]{Belokurov2008}
{Belokurov} V. {et~al.}, 2008, \apjl, 686, L83

\bibitem[{{Belokurov} {et~al}\mbox{.}(2010){Belokurov}, {Walker}, {Evans},
  {Gilmore}, {Irwin}, {Just}, {Koposov}, {Mateo}, {Olszewski}, {Watkins}, \&
  {Wyrzykowski}}]{Belokurov2010}
{Belokurov} V. {et~al.}, 2010, \apjl, 712, L103

\bibitem[{{Belokurov} {et~al}\mbox{.}(2009){Belokurov}, {Walker}, {Evans},
  {Gilmore}, {Irwin}, {Mateo}, {Mayer}, {Olszewski}, {Bechtold}, \&
  {Pickering}}]{Belokurov2009}
{Belokurov} V. {et~al.}, 2009, \mnras, 397, 1748

\bibitem[{Belokurov {et~al}\mbox{.}(2007)Belokurov, Zucker, Evans, Kleyna,
  Koposov, Hodgkin, Irwin, Gilmore, Wilkinson, Fellhauer, Bramich, Hewett,
  Vidrih, de~Jong, Smith, Rix, Bell, Wyse, Newberg, Mayeur, Yanny, Rockosi,
  Gnedin, Schneider, Beers, Barentine, Brewington, Brinkmann, Harvanek,
  Kleinman, Krzesinski, Long, Nitta, \& Snedden}]{Belokurov2007}
Belokurov V. {et~al.}, 2007, The Astrophysical Journal, 654, 897

\bibitem[{{Belokurov} {et~al}\mbox{.}(2006){Belokurov}, {Zucker}, {Evans},
  {Wilkinson}, {Irwin}, {Hodgkin}, {Bramich}, {Irwin}, {Gilmore}, {Willman},
  {Vidrih}, {Newberg}, {Wyse}, {Fellhauer}, {Hewett}, {Cole}, {Bell}, {Beers},
  {Rockosi}, {Yanny}, {Grebel}, {Schneider}, {Lupton}, {Barentine},
  {Brewington}, {Brinkmann}, {Harvanek}, {Kleinman}, {Krzesinski}, {Long},
  {Nitta}, {Smith}, \& {Snedden}}]{Belokurov2006}
{Belokurov} V. {et~al.}, 2006, \apjl, 647, L111

\bibitem[{{Brown}, {Vel{\'a}zquez} \& {Aguilar}(2005){Brown}, {Vel{\'a}zquez},
  \& {Aguilar}}]{Brown2005}
{Brown} A.~G.~A., {Vel{\'a}zquez} H.~M., {Aguilar} L.~A., 2005, \mnras, 359,
  1287

\bibitem[{{Brown} {et~al}\mbox{.}(2014){Brown}, {Tumlinson}, {Geha}, {Simon},
  {Vargas}, {VandenBerg}, {Kirby}, {Kalirai}, {Avila}, {Gennaro}, {Ferguson},
  {Mu{\~n}oz}, {Guhathakurta}, \& {Renzini}}]{Brown2014}
{Brown} T.~M. {et~al.}, 2014, \apj, 796, 91

\bibitem[{{Bullock}(2010)}]{Bullock2010}
{Bullock} J.~S., 2010, ArXiv e-prints

\bibitem[{{Chabrier}(2003)}]{Chabrier2003}
{Chabrier} G., 2003, \pasp, 115, 763

\bibitem[{{Close} {et~al}\mbox{.}(2003){Close}, {Siegler}, {Freed}, \&
  {Biller}}]{Close2003}
{Close} L.~M., {Siegler} N., {Freed} M., {Biller} B., 2003, \apj, 587, 407

\bibitem[{{de Bruijne}(2012)}]{deBruijne2012}
{de Bruijne} J.~H.~J., 2012, \apss, 341, 31

\bibitem[{{de Bruijne} {et~al}\mbox{.}(2015){de Bruijne}, {Allen}, {Azaz},
  {Krone-Martins}, {Prod'homme}, \& {Hestroffer}}]{deBruijne2015b}
{de Bruijne} J.~H.~J., {Allen} M., {Azaz} S., {Krone-Martins} A., {Prod'homme}
  T., {Hestroffer} D., 2015, ArXiv e-prints

\bibitem[{{de Bruijne}, {Rygl} \& {Antoja}(2015){de Bruijne}, {Rygl}, \&
  {Antoja}}]{deBruijne15}
{de Bruijne} J.~H.~J., {Rygl} K.~L.~J., {Antoja} T., 2015, ArXiv e-prints

\bibitem[{{de Souza} {et~al}\mbox{.}(2014){de Souza}, {Krone-Martins}, {dos
  Anjos}, {Ducourant}, \& {Teixeira}}]{DeSouza2014}
{de Souza} R.~E., {Krone-Martins} A., {dos Anjos} S., {Ducourant} C.,
  {Teixeira} R., 2014, \aap, 568, A124

\bibitem[{{Diehl} {et~al}\mbox{.}(2014){Diehl}, {Abbott}, {Annis}, {Armstrong},
  {Baruah}, {Bermeo}, {Bernstein}, {Beynon}, {Bruderer}, {Buckley-Geer},
  {Campbell}, {Capozzi}, {Carter}, {Casas}, {Clerkin}, {Covarrubias}, {Cuhna},
  {D'Andrea}, {da Costa}, {Das}, {DePoy}, {Dietrich}, {Drlica-Wagner},
  {Elliott}, {Eifler}, {Estrada}, {Etherington}, {Flaugher}, {Frieman}, {Fausti
  Neto}, {Gelman}, {Gerdes}, {Gruen}, {Gruendl}, {Hao}, {Head}, {Helsby},
  {Hoffman}, {Honscheid}, {James}, {Johnson}, {Kacprzac}, {Katsaros},
  {Kennedy}, {Kent}, {Kessler}, {Kim}, {Krause}, {Kron}, {Kuhlmann}, {Kunder},
  {Li}, {Lin}, {Maccrann}, {March}, {Marshall}, {Neilsen}, {Nugent}, {Martini},
  {Melchior}, {Menanteau}, {Nichol}, {Nord}, {Ogando}, {Old}, {Papadopoulos},
  {Patton}, {Petravick}, {Plazas}, {Poulton}, {Pujol}, {Reil}, {Rigby},
  {Romer}, {Roodman}, {Rooney}, {Sanchez Alvaro}, {Serrano}, {Sheldon},
  {Smith}, {Smith}, {Soares-Santos}, {Soumagnac}, {Spinka}, {Suchyta},
  {Tucker}, {Walker}, {Wester}, {Wiesner}, {Wilcox}, {Williams}, {Yanny}, \&
  {Zhang}}]{Diehl2014}
{Diehl} H.~T. {et~al.}, 2014, in Society of Photo-Optical Instrumentation
  Engineers (SPIE) Conference Series, Vol. 9149, Society of Photo-Optical
  Instrumentation Engineers (SPIE) Conference Series, p.~0

\bibitem[{{Duquennoy} \& {Mayor}(1991)}]{Duquennoy1991}
{Duquennoy} A., {Mayor} M., 1991, \aap, 248, 485

\bibitem[{{ESA}(1997)}]{ESA1997}
{ESA}, ed., 1997, ESA Special Publication, Vol. 1200, {The HIPPARCOS and TYCHO
  catalogues. Astrometric and photometric star catalogues derived from the ESA
  HIPPARCOS Space Astrometry Mission}

\bibitem[{{Forbes} \& {Kroupa}(2011)}]{Forbes2011}
{Forbes} D.~A., {Kroupa} P., 2011, PASA, 28, 77

\bibitem[{{Hern{\'a}ndez-P{\'e}rez} \& {Bruzual}(2013)}]{Hernandez-Perez2013}
{Hern{\'a}ndez-P{\'e}rez} F., {Bruzual} G., 2013, \mnras, 431, 2612

\bibitem[{{Holl}, {Hobbs} \& {Lindegren}(2010){Holl}, {Hobbs}, \&
  {Lindegren}}]{Holl2010}
{Holl} B., {Hobbs} D., {Lindegren} L., 2010, in IAU Symposium, Vol. 261, IAU
  Symposium, {Klioner} S.~A., {Seidelmann} P.~K., {Soffel} M.~H., eds., pp.
  320--324

\bibitem[{{Holl} \& {Lindegren}(2012)}]{Holl2012a}
{Holl} B., {Lindegren} L., 2012, \aap, 543, A14

\bibitem[{{Holl}, {Lindegren} \& {Hobbs}(2012){Holl}, {Lindegren}, \&
  {Hobbs}}]{Holl2012b}
{Holl} B., {Lindegren} L., {Hobbs} D., 2012, \aap, 543, A15

\bibitem[{{Hurley}, {Tout} \& {Pols}(2002){Hurley}, {Tout}, \&
  {Pols}}]{Hurley2002}
{Hurley} J.~R., {Tout} C.~A., {Pols} O.~R., 2002, \mnras, 329, 897

\bibitem[{Jordi {et~al}\mbox{.}(2010)Jordi, Gebran, Carrasco, de~Bruijne, Voss,
  Fabricius, Knude, Vallenari, Kohley, \& Mora}]{Jordi2010}
Jordi C. {et~al.}, 2010, A{\&}A, 523, A48

\bibitem[{{Kaiser} {et~al}\mbox{.}(2010){Kaiser}, {Burgett}, {Chambers},
  {Denneau}, {Heasley}, {Jedicke}, {Magnier}, {Morgan}, {Onaka}, \&
  {Tonry}}]{Kaiser2010}
{Kaiser} N. {et~al.}, 2010, in Society of Photo-Optical Instrumentation
  Engineers (SPIE) Conference Series, Vol. 7733, Society of Photo-Optical
  Instrumentation Engineers (SPIE) Conference Series, p.~0

\bibitem[{{Kirby} {et~al}\mbox{.}(2008){Kirby}, {Simon}, {Geha},
  {Guhathakurta}, \& {Frebel}}]{Kirby2008}
{Kirby} E.~N., {Simon} J.~D., {Geha} M., {Guhathakurta} P., {Frebel} A., 2008,
  \apjl, 685, L43

\bibitem[{{Klypin} {et~al}\mbox{.}(1999){Klypin}, {Kravtsov}, {Valenzuela}, \&
  {Prada}}]{Klypin1999}
{Klypin} A., {Kravtsov} A.~V., {Valenzuela} O., {Prada} F., 1999, \apj, 522, 82

\bibitem[{{Koposov} {et~al}\mbox{.}(2008){Koposov}, {Belokurov}, {Evans},
  {Hewett}, {Irwin}, {Gilmore}, {Zucker}, {Rix}, {Fellhauer}, {Bell}, \&
  {Glushkova}}]{Koposov08}
{Koposov} S. {et~al.}, 2008, \apj, 686, 279

\bibitem[{{Koposov} {et~al}\mbox{.}(2015){Koposov}, {Belokurov}, {Torrealba},
  \& {Wyn Evans}}]{Koposov2015}
{Koposov} S.~E., {Belokurov} V., {Torrealba} G., {Wyn Evans} N., 2015, ArXiv,
  astroph/1503.02079

\bibitem[{{Lada}(2006)}]{Lada2006}
{Lada} C.~J., 2006, \apjl, 640, L63

\bibitem[{{Laevens} {et~al}\mbox{.}(2015){Laevens}, {Martin}, {Ibata}, {Rix},
  {Bernard}, {Bell}, {Sesar}, {Ferguson}, {Schlafly}, {Slater}, {Burgett},
  {Chambers}, {Flewelling}, {Hodapp}, {Kaiser}, {Kudritzki}, {Lupton},
  {Magnier}, {Metcalfe}, {Morgan}, {Price}, {Tonry}, {Wainscoat}, \&
  {Waters}}]{Laevens2015}
{Laevens} B.~P.~M. {et~al.}, 2015, \apjl, 802, L18

\bibitem[{{Lindegren} {et~al}\mbox{.}(2012){Lindegren}, {Lammers}, {Hobbs},
  {O'Mullane}, {Bastian}, \& {Hern{\'a}ndez}}]{Lindegren2012}
{Lindegren} L., {Lammers} U., {Hobbs} D., {O'Mullane} W., {Bastian} U.,
  {Hern{\'a}ndez} J., 2012, \aap, 538, A78

\bibitem[{{Mateu} {et~al}\mbox{.}(2011){Mateu}, {Bruzual}, {Aguilar}, {Brown},
  {Valenzuela}, {Carigi}, {Vel{\'a}zquez}, \& {Hern{\'a}ndez}}]{Mateu2011}
{Mateu} C., {Bruzual} G., {Aguilar} L., {Brown} A.~G.~A., {Valenzuela} O.,
  {Carigi} L., {Vel{\'a}zquez} H., {Hern{\'a}ndez} F., 2011, \mnras, 415, 214

\bibitem[{{McConnachie}(2012)}]{McConnachie2012}
{McConnachie} A.~W., 2012, \aj, 144, 4

\bibitem[{{Mignard} {et~al}\mbox{.}(2008){Mignard}, {Bailer-Jones}, {Bastian},
  {Drimmel}, {Eyer}, {Katz}, {van Leeuwen}, {Luri}, {O'Mullane}, {Passot},
  {Pourbaix}, \& {Prusti}}]{Mignard08}
{Mignard} F. {et~al.}, 2008, in IAU Symposium, Vol. 248, IAU Symposium, {Jin}
  W.~J., {Platais} I., {Perryman} M.~A.~C., eds., pp. 224--230

\bibitem[{{Moore} {et~al}\mbox{.}(1999){Moore}, {Ghigna}, {Governato}, {Lake},
  {Quinn}, {Stadel}, \& {Tozzi}}]{Moore1999}
{Moore} B., {Ghigna} S., {Governato} F., {Lake} G., {Quinn} T., {Stadel} J.,
  {Tozzi} P., 1999, \apjl, 524, L19

\bibitem[{{Onions} {et~al}\mbox{.}(2012){Onions}, {Knebe}, {Pearce}, {Muldrew},
  {Lux}, {Knollmann}, {Ascasibar}, {Behroozi}, {Elahi}, {Han}, {Maciejewski},
  {Merch{\'a}n}, {Neyrinck}, {Ruiz}, {Sgr{\'o}}, {Springel}, \&
  {Tweed}}]{Onions2012}
{Onions} J. {et~al.}, 2012, \mnras, 423, 1200

\bibitem[{{Perryman} {et~al}\mbox{.}(2001){Perryman}, {de Boer}, {Gilmore},
  {H{\o}g}, {Lattanzi}, {Lindegren}, {Luri}, {Mignard}, {Pace}, \& {de
  Zeeuw}}]{Perryman2001}
{Perryman} M.~A.~C. {et~al.}, 2001, \aap, 369, 339

\bibitem[{{Press} \& {Schechter}(1974)}]{Press1974}
{Press} W.~H., {Schechter} P., 1974, \apj, 187, 425

\bibitem[{{Robin} {et~al}\mbox{.}(2012){Robin}, {Luri}, {Reyl{\'e}}, {Isasi},
  {Grux}, {Blanco-Cuaresma}, {Arenou}, {Babusiaux}, {Belcheva}, {Drimmel},
  {Jordi}, {Krone-Martins}, {Masana}, {Mauduit}, {Mignard}, {Mowlavi},
  {Rocca-Volmerange}, {Sartoretti}, {Slezak}, \& {Sozzetti}}]{Robin2012}
{Robin} A.~C. {et~al.}, 2012, \aap, 543, A100

\bibitem[{{Romero-G{\'o}mez} {et~al}\mbox{.}(2015){Romero-G{\'o}mez},
  {Figueras}, {Antoja}, {Abedi}, \& {Aguilar}}]{RomeroGomez15}
{Romero-G{\'o}mez} M., {Figueras} F., {Antoja} T., {Abedi} H., {Aguilar} L.,
  2015, \mnras, 447, 218

\bibitem[{{Shanks} {et~al}\mbox{.}(2013){Shanks}, {Belokurov}, {Chehade},
  {Croom}, {Findlay}, {Gonzalez-Solares}, {Irwin}, {Koposov}, {Mann},
  {Metcalfe}, {Murphy}, {Norberg}, {Read}, {Sutorius}, \&
  {Worseck}}]{Shanks2013}
{Shanks} T. {et~al.}, 2013, The Messenger, 154, 38

\bibitem[{{Simon} \& {Geha}(2007)}]{Simon2007}
{Simon} J.~D., {Geha} M., 2007, \apj, 670, 313

\bibitem[{{Springel}, {Frenk} \& {White}(2006){Springel}, {Frenk}, \&
  {White}}]{Springel2006}
{Springel} V., {Frenk} C.~S., {White} S.~D.~M., 2006, \nat, 440, 1137

\bibitem[{{Starck} \& {Murtagh}(2002)}]{Starck02}
{Starck} J.-L., {Murtagh} F., 2002, {Astronomical image and data analysis},
  {Starck, J.-L.~\& Murtagh, F.}, ed.

\bibitem[{{Tollerud} {et~al}\mbox{.}(2011){Tollerud}, {Bullock}, {Graves}, \&
  {Wolf}}]{Tollerud2011}
{Tollerud} E.~J., {Bullock} J.~S., {Graves} G.~J., {Wolf} J., 2011, \apj, 726,
  108

\bibitem[{{White} \& {Rees}(1978)}]{White1978}
{White} S.~D.~M., {Rees} M.~J., 1978, \mnras, 183, 341

\bibitem[{{Willman} {et~al}\mbox{.}(2005{\natexlab{a}}){Willman}, {Blanton},
  {West}, {Dalcanton}, {Hogg}, {Schneider}, {Wherry}, {Yanny}, \&
  {Brinkmann}}]{Willman2005b}
{Willman} B. {et~al.}, 2005{\natexlab{a}}, \aj, 129, 2692

\bibitem[{{Willman} {et~al}\mbox{.}(2002){Willman}, {Dalcanton}, {Ivezi{\'c}},
  {Jackson}, {Lupton}, {Brinkmann}, {Hennessy}, \& {Hindsley}}]{willman02}
{Willman} B., {Dalcanton} J., {Ivezi{\'c}} {\v Z}., {Jackson} T., {Lupton} R.,
  {Brinkmann} J., {Hennessy} G., {Hindsley} R., 2002, \aj, 123, 848

\bibitem[{{Willman} {et~al}\mbox{.}(2005{\natexlab{b}}){Willman}, {Dalcanton},
  {Martinez-Delgado}, {West}, {Blanton}, {Hogg}, {Barentine}, {Brewington},
  {Harvanek}, {Kleinman}, {Krzesinski}, {Long}, {Neilsen}, {Nitta}, \&
  {Snedden}}]{Willman2005a}
{Willman} B. {et~al.}, 2005{\natexlab{b}}, \apjl, 626, L85

\bibitem[{{Zucker} {et~al}\mbox{.}(2006{\natexlab{a}}){Zucker}, {Belokurov},
  {Evans}, {Kleyna}, {Irwin}, {Wilkinson}, {Fellhauer}, {Bramich}, {Gilmore},
  {Newberg}, {Yanny}, {Smith}, {Hewett}, {Bell}, {Rix}, {Gnedin}, {Vidrih},
  {Wyse}, {Willman}, {Grebel}, {Schneider}, {Beers}, {Kniazev}, {Barentine},
  {Brewington}, {Brinkmann}, {Harvanek}, {Kleinman}, {Krzesinski}, {Long},
  {Nitta}, \& {Snedden}}]{Zucker2006a}
{Zucker} D.~B. {et~al.}, 2006{\natexlab{a}}, \apjl, 650, L41

\bibitem[{{Zucker} {et~al}\mbox{.}(2006{\natexlab{b}}){Zucker}, {Belokurov},
  {Evans}, {Wilkinson}, {Irwin}, {Sivarani}, {Hodgkin}, {Bramich}, {Irwin},
  {Gilmore}, {Willman}, {Vidrih}, {Fellhauer}, {Hewett}, {Beers}, {Bell},
  {Grebel}, {Schneider}, {Newberg}, {Wyse}, {Rockosi}, {Yanny}, {Lupton},
  {Smith}, {Barentine}, {Brewington}, {Brinkmann}, {Harvanek}, {Kleinman},
  {Krzesinski}, {Long}, {Nitta}, \& {Snedden}}]{Zucker2006b}
{Zucker} D.~B. {et~al.}, 2006{\natexlab{b}}, \apjl, 643, L103

\end{thebibliography}
\IfFileExists{\jobname.bbl}{}
{\typeout{}
\typeout{****************************************************}
\typeout{****************************************************}
\typeout{** Please run "bibtex \jobname" to obtain}
\typeout{** the bibliography and then re-run LaTeX}
\typeout{** twice to fix the references!}
\typeout{****************************************************}
\typeout{****************************************************}
\typeout{}
}

\bsp

%\begin{thebibliography}{}
%\bibitem[Chandrasekhar(1943)]{ch43} Chandrasekhar, S.\ 1943, ApJ, 97, 255
%\end{thebibliography}
%\label{lastpage}

%\begin{thebibliography}{99}
%\bibitem[Chandrasekhar(1943)]{ch43} Chandrasekhar, S.\ 1943, ApJ, 97, 255
%\end{thebibliography}
\label{lastpage}

\end{document}